\documentclass[10pt,journal,compsoc]{IEEEtran}

\usepackage[nocompress]{cite}
\usepackage{lipsum}
\usepackage{graphicx}
\usepackage[caption=false, font=normalsize, labelfont=sf, textfont=sf]{subfig}
\usepackage{footnote}
\usepackage[ruled, linesnumbered]{algorithm2e}
\usepackage{xspace}
\usepackage{booktabs}
\usepackage{enumitem}
\usepackage{multirow}
\usepackage[normalem]{ulem}
\usepackage{amsmath}
\usepackage[export]{adjustbox}

\setlist{leftmargin=5.5mm}

\DeclareMathOperator*{\argmax}{arg\,max}
\newcommand{\ie}{\emph{i.e.,}\xspace}
\newcommand{\eg}{\emph{e.g.,}\xspace}
\newcommand{\etal}{\emph{et al.}\xspace}
\newcommand{\paratitle}[1]{\vspace{1ex}\noindent \textbf{#1}}

\newcommand{\st}{\tau}
\useunder{\uline}{\ul}{}

\begin{document}
\title{Pair-Linking for Collective Entity Disambiguation: Two Could Be Better Than All}

\author{Minh~C.~Phan,
	Aixin~Sun,
	Yi~Tay,
	Jialong~Han,
	and~Chenliang~Li%
	\thanks {This paper is an extended version of the CIKM conference paper~\cite{minh2017neupl}.}%
	\IEEEcompsocitemizethanks{
		\IEEEcompsocthanksitem M. C. Phan, A. Sun, Y. Tay and J. Han are with School of Computer Science and Engineering, Nanyang Technological University, Singpaore.\protect\\
		E-mail: phan0050@e.ntu.edu.sg; axsun@ntu.edu.sg;\protect\\
		ytay017@e.ntu.edu.sg; jialonghan@gmail.com
		\IEEEcompsocthanksitem 	C.  Li  is  with School of Cyber Science and Engineering, Wuhan University, China.\protect\\
		E-mail: cllee@whu.edu.cn%
	
	}

}
\markboth{IEEE TRANSACTIONS ON KNOWLEDGE AND DATA ENGINEERING, Accepted in 2018}%
{}

\IEEEtitleabstractindextext{
\begin{abstract}
Collective entity disambiguation, or collective entity linking aims to jointly resolve multiple mentions by linking them to their associated entities in a knowledge base. Previous works are primarily based on the underlying assumption that entities within the same document are highly related. However, the extent to which these entities are actually connected in reality is rarely studied and therefore raises interesting research questions. For the first time, this paper shows that the semantic relationships between mentioned entities within a document are in fact less dense than expected. This could be attributed to several reasons such as noise, data sparsity, and knowledge base incompleteness. As a remedy, we introduce MINTREE, a new tree-based objective for the problem of entity disambiguation. The key intuition behind MINTREE is the concept of \textit{coherence relaxation} which utilizes the weight of a minimum spanning tree to measure the coherence between entities. Based on this new objective, we design Pair-Linking, a novel iterative solution for the MINTREE optimization problem. The idea of Pair-Linking is simple: instead of considering all the given mentions, Pair-Linking iteratively selects a pair with the highest confidence at each step for decision making. Via extensive experiments on 8 benchmark datasets, we show that our approach is not only more accurate but also surprisingly faster than many state-of-the-art collective linking algorithms.

\end{abstract}
	
}

\maketitle

\IEEEraisesectionheading{\section{Introduction}\label{sec:introduction}}

\IEEEPARstart{M}{entions} of named entities such as people, places, and organizations are commonplace in documents. However, these mentions of named entities are usually ambiguous due to the polymorphic nature of language \ie the same entity may be mentioned in different surface forms, and the same surface form may refer to different named entities. Entity disambiguation alleviates this problem by bridging unstructured text and structured knowledge bases \ie named entities are effectively disambiguated by a knowledge base assignment. The desirable benefit of entity disambiguation enables and benefits a myriad of downstream practical applications which include knowledge base population, information retrieval and extraction, question answering, and content analysis. As such, it has received considerable attention across both industrial and academic research communities. 

The problem of entity disambiguation can be described as follows: Given a document containing a set of mentions, the task is to assign each mention to a correct entity in a provided knowledge base. Take the following sentence from Wikipedia as an example:

\vspace{0.3cm}
\textit{``Before turning seven, \underline{Tiger} won the Under Age 10 section of the Drive, Pitch, and Putt competition, held at the Navy Golf Course in Cypress, California".}
\vspace{0.3cm}

Without considering its context, the word `Tiger' can refer to the American golfer \textit{Tiger Woods}, the budget airline \textit{Tiger Air}, or the beer brand \textit{Tiger Beer}. When context is taken into account, the mention `Tiger' in the given sentence should be linked to golfer \textit{Tiger\_Woods}\footnote{https://en.wikipedia.org/wiki/Tiger\_Woods}.

\paratitle{The Research Problem.} Formally, given a document $d$ containing a set of mentions $M = \{m_1,...,m_N\}$ and a target knowledge base $W$, the task of entity disambiguation is to find a mapping $M \mapsto W$ that links each mention to a correct entity in the knowledge base.
We denote the output of the matching as an $N$-tuple $\Gamma = (e_1,...,e_N)$ where $e_i$ is the assigned entity for mention $m_i$ and $e_i \in W$. Similar to most recent works~\cite{PappuBMST17, GaneaGLEH16, yamada2016joint}, we do not address the issue of mention extraction and not-in-list identification in this study. That is, every mention has a corresponding entity in the given knowledge base.

\begin{figure}
	\centering
	\subfloat [Example 1
	\label{fig:corr_eg1}]{\fbox{\includegraphics[scale=0.5,trim=2 2 2 2,clip]{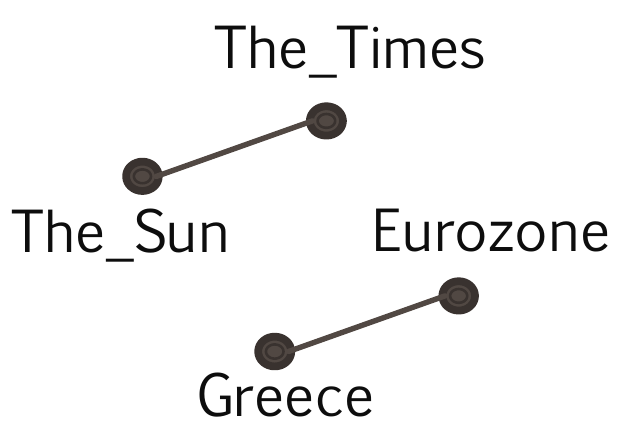}}}
	\hspace{0.5cm}
	\vspace{0.5cm}		
	\subfloat[Example 2
	\label{fig:corr_eg2}]{\fbox{\includegraphics[scale=0.5,trim=2 2 2 2,clip]{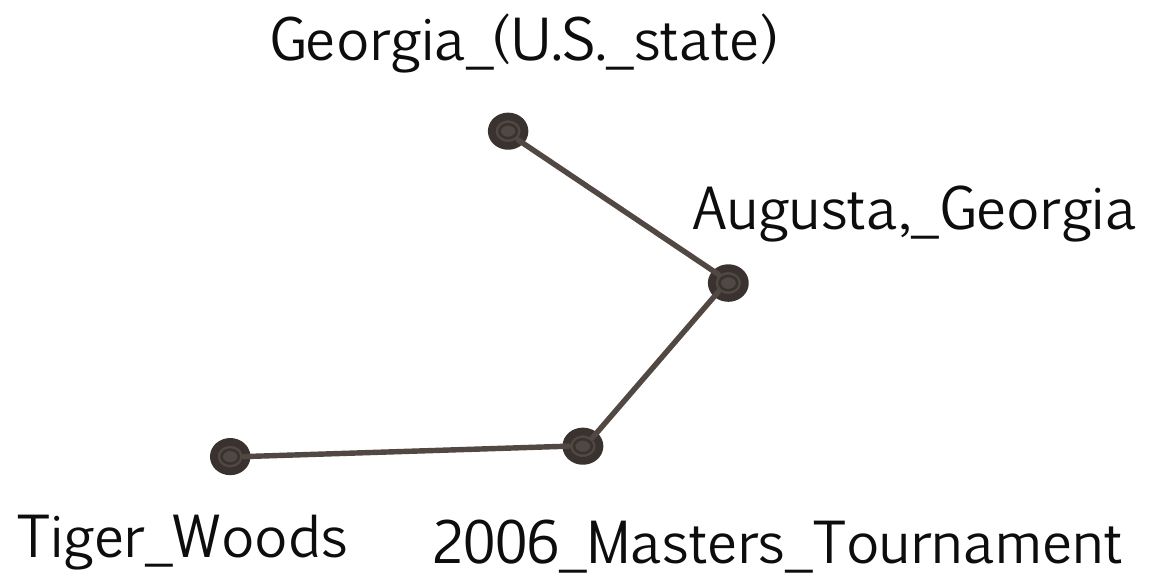}}}
	\caption{Illustration of coherence between linked entities in two examples. The edges represent strong semantic relatedness between entities.}
\end{figure}

While many early works in named entity disambiguation solely rely on local contexts of the entity mentions, our work is mainly concerned with utilizing the \textit{coherence} between linked entities (in the document) to improve the disambiguation. This is known as \textit{collective} entity disambiguation and it is widely adopted amongst many competitive approaches. In many of these approaches, the basic intuition is that each linked entity should obey and maintain a maximal `pairwise' coherence with all other entities. We refer to these approaches as \textbf{ALL-Link}.

ALL-Link is based on the assumption that all entities mentioned in a document are densely connected in the knowledge base (KB). Unfortunately, it is easy to see that this is not always the case, \ie not all entities mentioned in a document are always highly related to each other. Consider the following examples:

\begin{enumerate}
	\item ``\underline{The Sun} and \underline{The Times} reported that \underline{Greece} will have to leave the \underline{Euro} soon''.
	\item ``\underline{Wood} played at \underline{2006 Master} held in \underline{Augusta}, \underline{Georgia}''
\end{enumerate}
where entity mentions are underlined. In the first example, only two entities are closely related, which is shown in Figure~\ref{fig:corr_eg1}. On the other hand, the entities in Figure~\ref{fig:corr_eg2} are connected in a chain-like form. Both examples illustrate the sparse coherence (between mentioned entities) which is commonplace in generic documents. This qualitatively shows that the fundamental assumption and objective of ALL-Link leaves much to be desired. 

In lieu of the apparent weakness of ALL-Link, this paper proposes a novel and simple paradigm. Our approach relaxes the pairwise coherence assumption and affirms the narrative that maintaining pairwise coherence between all entities is \textbf{unnecessary}. Furthermore, relaxation of this assumption allows us to significantly improve not only the accuracy but also the runtime of collective entity disambiguation. Overall, the prime contributions of this work are as follows:

\begin{itemize}
	\item For the first time, we study the form of coherence between mentioned entities (\ie whether it is sparse or dense). We show that not all entities (in a general document) are highly related to each other. This insight leads us to develop a new objective that relaxes the coherence condition, aiming towards an more effective and faster solution for entity disambiguation.
	
	\item We propose a tree-based model that utilizes the weight of spanning tree as the linking objective. We provide a detailed analysis showing that our proposed tree-based objective is highly correlated with the conventional objectives and it can be used to effectively model the disambiguation quality.
	
	\item We introduce Pair-Linking, an approximate solution for the tree-based model. Pair-Linking achieves state-of-the-art performance while being extremely fast in comparison to other collective linking algorithms. 
	
\end{itemize}

\section{Related Work}
\label{sec:related}

\begin{figure*}
	\centering
	\includegraphics[scale=0.45,]{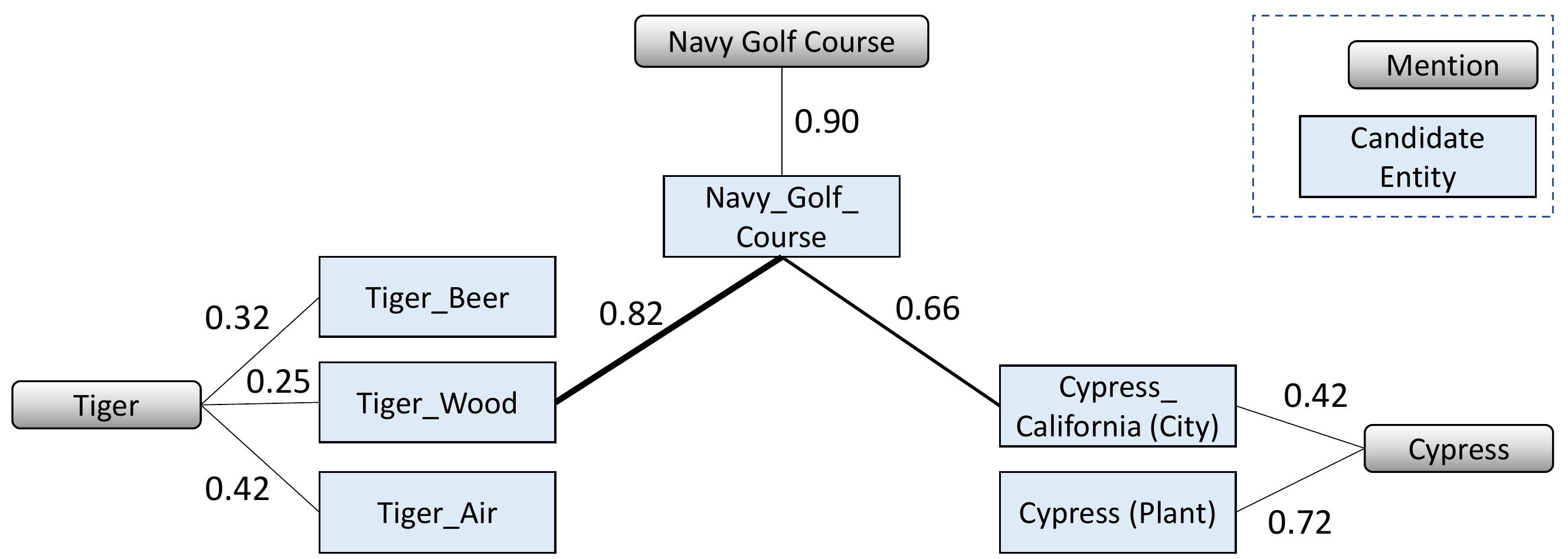}
	\caption{An example of mention-entity graph. The weights between mentions and entities represent the local confidence while the weights between entities represents the semantic relatedness.}
	\label{fig:mention_entity_graph}
\end{figure*}

Collective entity disambiguation approaches can be characteristically dichotomized into two families: optimization-based approach and graph-based approach. The optimization-based approach models the entity disambiguation problem and solves the objective function by optimization techniques. On the other hand, graph-based approach directly approximates the solution by doing influence propagation on the mention-entity graph built from the mentions and candidate entities. We will describe the two approaches in this section.

\subsection{Optimization-based Approach}
The common technique for finding the optimal disambiguation, denoted by $\Gamma^*$, is to maximize the local confidence of each assignment $ \phi(m_i,e_i)$, while enforcing the pairwise coherence among all linked entities $\psi(e_i,e_j)$. The objective is referred to as \textbf{ALL-Link} and is expressed as follows:

\begin{equation}
\Gamma^* = \argmax_{\Gamma} \left[\sum_{i=1}^N \phi(m_i,e_i) + \sum_{i=1}^N \sum_{j=1, j\neq i}^N \psi(e_i,e_j)\right]\label{eq:all_link_objective}
\end{equation}

Local confidence or local score $\phi(m_i,e_i)$ reflects the likelihood of mappings $m_i \mapsto e_i$ based on $m_i$'s textual context and $e_i$'s profile, in regardless to other assignments in the same document. It is computed through the textual similarity between the entity mention and the candidate entity's name, and/or the semantic similarity between the context around the entity mention and the document associated with the candidate entity ~\cite{ShenWH15}. On the other hand, the pairwise coherence $\psi(e_i, e_j)$ represents the semantic relatedness between entities and it is often computed by utilizing the linkage structure in knowledge base (\eg Wikipedia) or entity embedding. Detailed computation of the two components will be described in Section~\ref{sec:preliminaries}.

The optimization expressed in Equation~\ref{eq:all_link_objective} is NP-Hard, therefore, Shen~\etal~\cite{ShenWLW12} use iterative substitution (\ie hill climbing technique) to approximate the solution: the optimal assignment is obtained by substituting an assignment $m_i \mapsto e_i$ with another mapping $m_i \mapsto e_j$ as long as it improves the objective score. In the other works \cite{GaneaGLEH16, GlobersonLCSRP16}, \textit{Loopy Belief Propagation} (LBP)~\cite{MurphyWJ99} is utilized. Both approaches have the complexity of $\mathcal{O} \left(\mathcal{I}\times N^2k^2\right)$ where $\mathcal{I}$ is the number of iterations required for convergence, $N$ and $k$ are the number of mentions and candidates per mention respectively.

Other methods follow the idea proposed by Ratinov~\etal~\cite{RatinovRDA11}. First, they extract a set of unambiguous mentions based on the local confidence score $\phi(m_i,e_i)$. The associated set of linked entities will be used as the disambiguation context $\Gamma'$. The global optimization task is then decomposed into the optimization of individual coherences, described by the formula:

\begin{equation}
\Gamma^* = \argmax_{\Gamma} \sum_{i=1}^N \left[\phi(m_i,e_i) + \sum_{e_j\in \Gamma'} \psi(e_i,e_j)\right]\label{eq:decomposed_obj_1}
\end{equation}

The challenge with the Ratinov's approach is that the unambiguous set of mention is not always obtainable beforehand. In many cases, all mentions within a document can be ambiguous due to the noisy and unclear context. Therefore, to disambiguate a mention, authors in \cite{FerraginaS10, GlobersonLCSRP16} consider the evidence from not only the unambiguous mentions but also the ambiguous ones. Specifically, suppose $S_{ij}(e_i)$ is the support for label $e_i$ from mention $m_j$, then $S_{ij}(e_i)$ is defined as follows:

\begin{equation}
S_{ij}(e_i) = \max_{e_j}\left[\phi(m_j,e_j) + \psi(e_i,e_j)\right]
\label{eq:decomposed_obj_2}
\end{equation}

The disambiguated entity $e_i$ for mention $m_i$ is extracted as follows:

\begin{equation}
e_i = \argmax_{e_i}\left[\phi(m_i,e_i) + \sum_{j=1,j\neq i}^{N}S_{ij}(e_i)\right]
\label{eq:decomposed_obj_2'}
\end{equation}

Interestingly, the work in~\cite{GlobersonLCSRP16} reveals that the best performance is obtained by considering evidence from not all but only \textit{top-k} supporting mentions. Furthermore, the authors also study the \textbf{SINGLE-Link}, which considers only the most related evidence. Its disambiguation objective is expressed as follows:

\begin{equation}
\Gamma^* = \argmax_{\Gamma} \sum_{i=1}^N  \left[\phi(m_i,e_i) + \max_{j=1}^N \psi(e_i,e_j)\right]\label{eq:sigle_link_objective}
\end{equation}

In another work~\cite{PappuBMST17}, fast collective linking is achieved by just looking at only the neighbouring connections \ie the previous and subsequent mentions. The associated objective function can be written as follows:

\begin{equation}
\Gamma^* = \argmax_{\Gamma} \left[\sum_{i=1}^N \phi(m_i,e_i) + \sum_{i=1}^{N-1} \psi(e_i,e_{i+1})\right]\label{eq:fwbw_objective}
\end{equation}

Dynamic programming, specifically Forward-Backward algorithm~\cite{austin1991forward} (FwBw) is utilized to solve the optimization above. Although this approach works well on short text (\ie query)~\cite{PappuBMST17}, it is incapable of capturing long-distance coherence which is important for disambiguation in long documents.

\subsection{Graph-based Approach}
Graph-based approaches solve the disambiguation problem by performing collective linking on mention-entity graph. The graph is constructed with edges connecting mentions and their candidate entities. The edges are weighted by the score of local context matching $\phi(m_i,e_i)$. There are also edges connecting between candidate entities. Their weights reflect the semantic coherence $\psi(e_i, e_j)$. An example of a mention-entity is illustrated in Figure~\ref{fig:mention_entity_graph}.

Hoffart~\etal~\cite{HoffartYBFPSTTW11} cast the joint mapping into the problem of identifying dense subgraph that contains exactly one candidate entity for each mention. Many other works are based on the Random Walk and PageRank algorithms~\cite{HanSZ11,HacheyRC11,GuoB14,PiccinnoF14,AlhelbawyG14,0001RN14}. Specifically, authors in \cite{ZwicklbauerSG16} introduce a new 'pseudo` topic node into the mention-entity graph to enforce the agreement between the disambiguated entities and the topic node's context. The node is initialized by all the unambiguous mentions. In DoSeR~\cite{ZwicklbauerSG16}, Personalized PageRank is iteratively performed on the mention-entity graph. At each step, candidate entities with high stabilized scores will be selected to map to its associated mentions and the entities are added into the pseudo topic node. Although graph-based approaches are shown to produce competitive performance, they are computationally expensive, especially in case of long documents containing hundreds of mentions.

\paratitle{Discussion.}
Existing studies on collective linking problem either propose an objective and its solution (\eg the optimization based approaches) or directly approximate the problem (\eg PageRank). There is no prior work that studies the coherence structure of the mentioned entities. Specifically, the research question is ``\textit{to what extent the mentioned entities are related to each other? (by a specific relatedness measure)}``. To the best of our knowledge, we are the first to address this research problem. We also study a new tree-based objective used to model the coherence between entities.

We acknowledge a portion of related works about name disambiguation in bibliographic databases~\cite{TangFWZ12, ZhangH17, CenDSO13}. Although the ideas introduced in these works can be transferable to entity disambiguation, utilization of the proposed techniques is not directly applicable to our problem setting as we focus on collective linking algorithms. 

\section{Preliminaries}
\label{sec:preliminaries}

\begin{table}
	\centering
	\caption{Frequently used notations.}
	\label{tb:notations}
	\begin{tabular}{p{1.2cm}|p{6.8cm}}
		\toprule
		\textbf{Notation}&  \textbf{Definition and description}\\
		\midrule
		$M$	& List of mentions to be linked in a document.\\
		$m_i$	& The $i^{th}$ mention in M.\\
		$W$	& Set of all entities in knowledge base (we use Wikipedia in this work).\\
		$e_i$	& An entity in  $W$ that is assigned to mention $m_i$.\\
		$N$	& Number of mentions in a document.\\
		$k$	& Number of candidate entities for each mention.\\
		$C_i$	& List of candidate entities for mention $m_i$.\\
		$e_i^k$	& The $k^{th}$ candidate entity of mention $m_i$.\\		
		$\Gamma$	& A mapping $M\mapsto W$ that represents a disambiguation result.\\
		$\phi(m_i,e_i)$	& Local confidence of mapping $m_i$ to $e_i$.\\
		$\psi(e_i,e_j)$	& Pairwise coherence or semantic relatedness between two entities, $e_i$ and $e_j$.\\
		$d(e_i,e_j)$	& Semantic distance between two entities, $e_i$ and $e_j$, in the MINTREE coherence graph.\\	
		\bottomrule
	\end{tabular}
\end{table}

In this section, we give an overview about concepts and components used in out disambiguation system. For the ease of presentation, we summarize primary notations used throughout this paper in Table~\ref{tb:notations}.

Given a document with a set of mentions to be disambiguated, the candidate entities for each mention are extracted based on the mention's surface form. Collective linking works on the sets of candidate entities and selects for each mention an entity that optimizes the objective consists of local confidence $\phi(m_i,e_i)$ and the pairwise coherence $\psi(e_i,e_j)$ (described in the previous section). 

It is worth mentioning that the work in this paper does not focus on improving the local confidence or the semantic relatedness. Our work mainly focuses on the study and evaluation of different collective linking models and solutions. In this section, we will detail the methods that are commonly used to compute the local confidence and the semantic relatedness. First, we will describe the word and entity embeddings that will be used in the later calculations.

\paratitle{Word and Entity Embeddings.}
Embedding models aim to generate a continuous representation for every word, such that two words that are close in meaning are also close in the embedding vector space. It assumes that words are similar if they co-occur often~\cite{mikolov2013distributed}. Correspondingly, we can assume two entities to be semantically related if they are found in an analogous context. The context is defined by the surrounding words or entities.

Jointly modeling of words and entities in the same continuous space has been shown to improve the quality of both word and entity embeddings~\cite{wang2014knowledge}, and benefits entity disambiguation task~\cite{yamada2016joint, fang2016entity}. In this work, we use the word2vec with skip-gram model~\cite{mikolov2013distributed} to jointly learn the distributional representation of words and entities.

Let $\mathcal{T}$ denote the set of tokens. Token $\tau \in \mathcal{T}$ can be either a word (\eg Tiger, Wood) or an $entityID$ (\eg \textit{[Tiger\_Wood]}). Given a sequence of tokens $ \st_1,...,\st_N$, the skip-gram model tries to maximize the following average log probability:

\begin{equation}
\mathcal {L} = \dfrac{1}{N} \sum_{i=1}^N {\sum_{-c \le j \le c,j \ne 0} \log{} P (\st_{i+j}|\st_i)}\label{eq:lossw2v}
\end{equation}
where $c$ is the size of context window, $\st_i$ denotes the target token, and $\st_{i+j}$ is a context token. The conditional probability $P (\st_{i+j}|\st_i)$ is defined by the softmax function:

\begin{equation}
P (\st_{O}|\st_I) = \dfrac	{\exp({v'_{\st_O}}^\mathsf{T} v_{\st_I})}
{\sum_{\tau \in \mathcal{T}} {\exp({v'_{\st}}^\mathsf{T} v_{\st_I})}}\label{eq:w2vprob}
\end{equation}
where $v_\st$ and $v'_\st$ are the `input' and `output' vector representations of $\st$, respectively. After training, we use the `output' $v'_\st$ as the embedding for word or entity.

To co-train word and entity embeddings, we create a `token corpus' by exploiting the existing hyperlinks in Wikipedia. Specifically, for each sentence in Wikipedia which contains at least one hyperlink to another Wikipedia entry, we create an additional sentence by replacing each anchor text with its associated entityID. For each Wikipedia page, we also create a `pseudo sentence' which is the sequence of entityIDs linked from this page, in the order of their appearances. For example, assume that the Wikipedia page about \textit{Tiger Wood} contains only 2 sentences: ``\textit{Woods}\textsubscript{[Tiger\_Woods]} was born in \textit{Cypress}\textsubscript{[Cypress,\_California]}. He has a niece, \textit{Cheyenne Woods}\textsubscript{[Cheyenne\_Woods]}.", the following sentences are included in our `token corpus'.
\begin{itemize}
	\item Wood was born in Cypress. He has a niece, Cheyenne Woods.
	\item \textit{[Tiger\_Woods]} was born in \textit{[Cypress,\_California]}. He has a niece, \textit{[Cheyenne\_Woods]}.	
	\item \textit{[Tiger\_Woods]} \textit{[Cypress,\_California]} \textit{[Cheyenne\_Woods]}.
\end{itemize}

\paratitle{Local Confidence Score $\phi(m_i,e_i)$.} We adopt the approach proposed in~\cite{yamada2016joint} to estimate the matching score between a mention (with its local context) and a candidate entity. Specifically, a learning to rank model, Gradient Boosting Tree, is trained to estimate the probability that a mention $m_i$ will be mapped to a candidate entity $e_i$. 

The features to be used include the prior probability that an entity is selected given the mention's surface form $P(e|m)$, several string similarity features between the mention's surface form and the entity's title, and finally the semantic similarity between the candidate entity and the mention's surrounding context. The raw output obtained from the ranking model  will be used as the local confidence score.

It is worth mentioning that there are more effective ways to estimate the local confidence with the use of deep neural networks~\cite{minh2017neupl, SunLTYJW15, Francis-LandauD16}. However, this is not the focus of this work and we will implement the most efficient way for the estimation (as described above).

\paratitle{Pairwise Coherence Score (or Relatedness Measure) $\psi(e_i,e_j)$.} We study a wide range of semantic similarity measures ($\psi(e_i,e_j)$) including the Wikipedia Links-based measure and the Entity Embedding similarity. The Wikipedia Link-based measure (WLM)~\cite{MilneW08} is widely used to estimate the coherence. It is based on the assumptions that two entities are related if there are many Wikipedia pages that link to both. The WLM score for two entities $e_1$, $e_2$ is calculated as follows:

\begin{equation}
\small
WLM(e_1,e_2) = 1 - \frac{\log(\max(|U_1|,|U_2|)+1) - \log(|U_1\cap U_2|+1)}{\log(|W|+1) - \log(\min(|U_1|,|U_2|)+1)}
\end{equation}
where $|U_1|$ and $|U_2|$ are the set of Wikipedia articles that have hyperlinks to $e_1$ and $e_2$ respectively, and $W$ is the set of all Wikipedia articles.

We also exploit Jaccard-like similarity. Different from the original formula in Guo~\etal~\cite{GuoCK13}, here we take logarithm scale as it yields better results. The Normalized Jaccard Similarity (NJS) is then defined as follows:

\begin{equation}
NJS(e_1,e_2) = \frac{\log(|U_1\cap U_2|+1)}{\log(|U_1\cup U_2|+1)}
\end{equation}

Furthermore, we study the entity embedding similarity (EES) which is the cosine similarity of the two representations:

\begin{equation}
EES(e_1,e_2) = cos(embeding(e_1),embeding(e_2))
\end{equation}

The embedding of entity is trained jointly with word's embedding taken from the Wikipedia corpus. Using the entity embedding to estimate the semantic relatedness has shown to be effective for entity disambiguation in recent works~\cite{ZwicklbauerSG16, yamada2016joint, PappuBMST17}.

\section{Entitative coherence in document}
\label{sec:coherenec_analysic}

\begin{figure}
	\centering
	\subfloat[Dense
	\label{fig:dense_form}]{\fbox{\includegraphics[scale=0.45,trim=-0 10 0 10,clip]{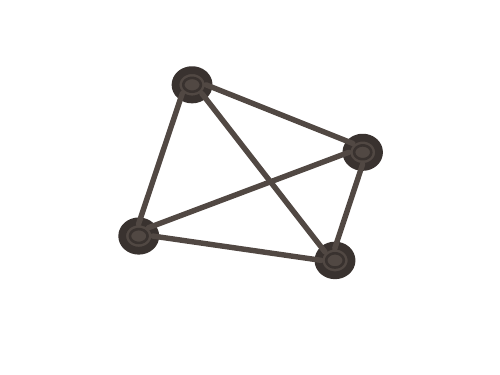}}}
	\hspace{1.0cm}
	\subfloat[Tree-like
	\label{fig:tree_form1}]{\fbox{\includegraphics[scale=0.45,trim=-0 10 0 10,clip]{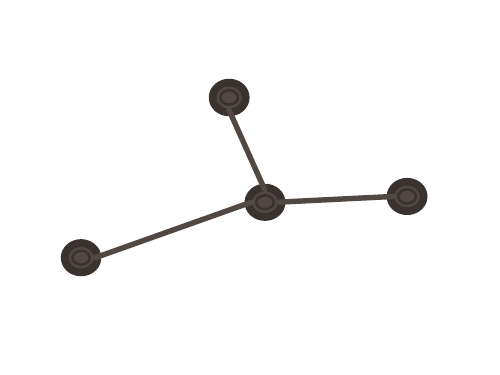}}}
	\hspace{2.5cm}
	\vspace{0.5cm}
	\subfloat[Chain-like
	\label{fig:tree_form2}]{\fbox{\includegraphics[scale=0.45,trim=-0 10 0 10,clip]{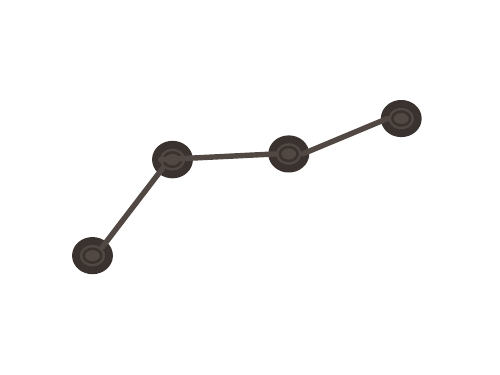}}}
	\hspace{1.0cm}
	\subfloat [Forest-like
	\label{fig:forest_form}]{\fbox{\includegraphics[scale=0.45,trim=-0 10 0 10,clip]{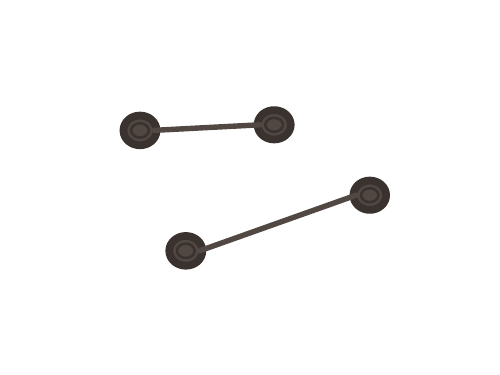}}}
	
	\caption{Illustration of four different levels of denseness in entity coherence graph.}
	\label{fig:4form}
\end{figure}

\begin{table*}
	\centering
	\caption{Average coherence denseness calculated for each dataset. Only documents whose number of mentions greater than 3 are considered. The results are reported with three relatedness measures: Wikipedia Link-based Measure (WLM), Normalized Jaccard Similarity (NJS) and Entity Embedding Similarity (EES).}
	\label{tb:coher_graph_degree}
	\begin{tabular}{l||c||ccc||ccc}
		\toprule
		\multirow{2}{*}{\textbf{Dataset}} & \multirow{2}{*}{\textbf{$|D|$}} & \multicolumn{3}{c}{\textbf{$Coh\_deg$ (theoretical)}} & \multicolumn{3}{c}{\textbf{$Coh\_deg$ (calculated)}} \\
		&                              & Forest        & Tree       & Dense       & WLM     & NJS    & EES    \\
		\midrule
		Reuters128	&30	&1.00	&1.64	&5.93	&3.21	&2.13	&2.68\\
		ACE2004	&25	&1.00	&1.69	&7.20	&3.23	&2.83	&2.75\\
		MSNBC	&19	&1.00	&1.83	&14.89	&6.35	&4.48	&7.08\\
		Dbpedia	&35	&1.00	&1.71	&6.60	&3.08	&2.55	&2.92\\
		KORE50	&9	&1.00	&1.54	&3.44	&1.36	&1.58	&1.36\\
		Micro14	&80	&1.00	&1.53	&3.33	&1.81	&1.72	&1.82\\
		AQUAINT	&50	&1.00	&1.84	&12.82	&5.78	&3.39	&4.53\\
		\bottomrule
	\end{tabular}
\end{table*}

As illustrated by two examples in the introduction section, documents (in general) may contain non-salient entities or entities that do not have complete connections in knowledge base. Therefore, the basic assumption used by conventional collective linking approaches (all the entities mentioned should be densely related) leaves much to be desired. For the first time, we study the form of coherence between the entities in a document. In this section, we will measure the denseness of entity connection in 8 testing datasets (details about each dataset will be presented in Section~\ref{ssec:datasets_methods}).

Suppose a graph $G(V,E)$ contains all the entities mentioned in a document. The edges between every pair of entities are weighted by the semantic relatedness. We will analyse and report the result with all three relatedness measures: the Wikipedia link-based measures (WLM), the normalized Jaccard similarity (NJS) and the entity embedding (cosine) similarity (EES).

Note that our intent is to measure the denseness (or sparseness) of the connections, not the degree of coherence. The degree of coherence can be estimated through the average weight of the relatedness graph. However, we are more interested in knowing whether the entities are densely or sparsely connected regardless of the coherence degree.

Figure~\ref{fig:4form} illustrates four standard forms of the coherence between entities. Focusing on the denseness, if all pairs of entities are connected at the same coherence degree (can be at high or low pairwise coherence score), we would say the entities are densely connected (Figure~\ref{fig:dense_form}). On the other hand, if there are only few pairs dominating the pairwise coherence, we will view it as sparse (Figures~\ref{fig:forest_form},~\ref{fig:tree_form1},~\ref{fig:tree_form2}).

The illustration hints that the denseness of a coherence graph can be estimated through the average degree of its filtered graph $G_\theta(V,E_\theta)$. The filtered graph $G_\theta$ consists of only the edges having highest pairwise relatedness scores (\ie $E_\theta=\{e|e\in E \wedge weight(e) \geq \theta \}$). The threshold $\theta$ needs to be carefully set for each entity graph. If the threshold is too high, a small number of edges will be left in the filtered graph, resulting in a low denseness score. On the other hand, if the value is small, the average degree of the filtered graph will be high. To this end, we determine a dynamic threshold $\theta$ for each document as follows. The $\theta$ is chosen as the largest value such that every vertex (or entity) in $V$ is incident to at least one edge in $E_\theta$. Intuitively, each coherence graph is pruned to the same `standard form' before calculating its average degree. In other words, the associated filtered edge set $E_\theta$ will be a valid \textit{edge cover}\footnote{https://en.wikipedia.org/wiki/Edge\_cover} of the graph G. Finally, we calculate the average degree of $G_\theta(V,E_\theta)$ and refer to it as the denseness of coherence for the entity set $V$.

\begin{equation}
Denseness(V) = Avg\_deg(G_\theta) = \frac{2\times|E_\theta|}{|V|}
\end{equation}

Note that the filtered graph $G_\theta$ contains highly related connections between entities. The average degree of $G_\theta$ will reflect the density of the connections. Higher value means that the entity set V is densely connected, and lower value indicates the sparse coherence among the entities. As illustrated in Figure~\ref{fig:forest_form}, if $G_\theta$ is sparse (\ie every entity is strongly related to only one other entity), its theoretical average degree is equal to $1$. On the other hand, if entities in $G_\theta$ are connected by tree-like or chain-like fashion (see Figures~\ref{fig:tree_form1},~\ref{fig:tree_form2}), the denseness value is $2*(n-1)/n$. Furthermore, the expected value for densely connected case (Figure~\ref{fig:dense_form}) is close to $(n-1)$  where n is the number of entities (or vertices).

We report the coherence denseness for 7 benchmark datasets in Table~\ref{tb:coher_graph_degree}. We consider only the documents having at least 4 mentions because the ones with 3 or fewer mentions will lead to a fixed denseness score by the calculation described above. It is also worth mentioning that for short text datasets like KORE50 or Micro14, the edge filtering is more likely to prune the entity graph into a tree-like form or forest-like form, leading to the bias in the denseness score. However, for completeness, we report the scores of both short and long text datasets in Table~\ref{tb:coher_graph_degree}. 

Table~\ref{tb:coher_graph_degree} shows, in general, the calculated values lie closer to the tree (or chain) form's expected values rather than the dense form. The same result is observed in all settings of relatedness measures (WLM, NJS, and EES). Especially, in long text datasets like MSNBC and AQUAINT, each mentioned entity is highly related to only 3-5 other entities (by the NJS measure) although the number of entities in each document in the two datasets is more than 13 (on average). The result reveals that not all entities mentioned in a document are densely related to each other; therefore, considering all the pairwise connections is \textbf{not necessary} for collective entity disambiguation.

Next, we define a new graph-based model that relaxes the ALL-Link coherence objective (Equation~\ref{eq:all_link_objective}), and allow us to propose a fast and effective linking algorithm.

\section{Minimum Spanning Tree Representative}
\label{sec:model}
We introduce MINTREE, a new tree-based objective to effectively model the entity disambiguation problem. First, we define a new coherence measure for a set of entities. 

\paratitle{MINTREE Coherence Measure.} Given a set of entities V and its associated entity relatedness graph $G(V,E)$, the edges connecting all pair of entities are weighted by a specific semantic distance. The coherence of the graph G is defined as the weight of the minimum-spanning tree (MST) that can be formed in G.

The MINTREE coherence measure defined in this way relaxes the conventional ALL-Link-like objective which is the sum of all edge's weights in G. Next, we formulate the collective entity disambiguation problem that utilizes the new measure as the objective to optimize.

\paratitle{MINTREE Problem Statement.}
\begin{figure}
	\centering
	\includegraphics[scale=0.65,,trim=2 2 2 2,clip]{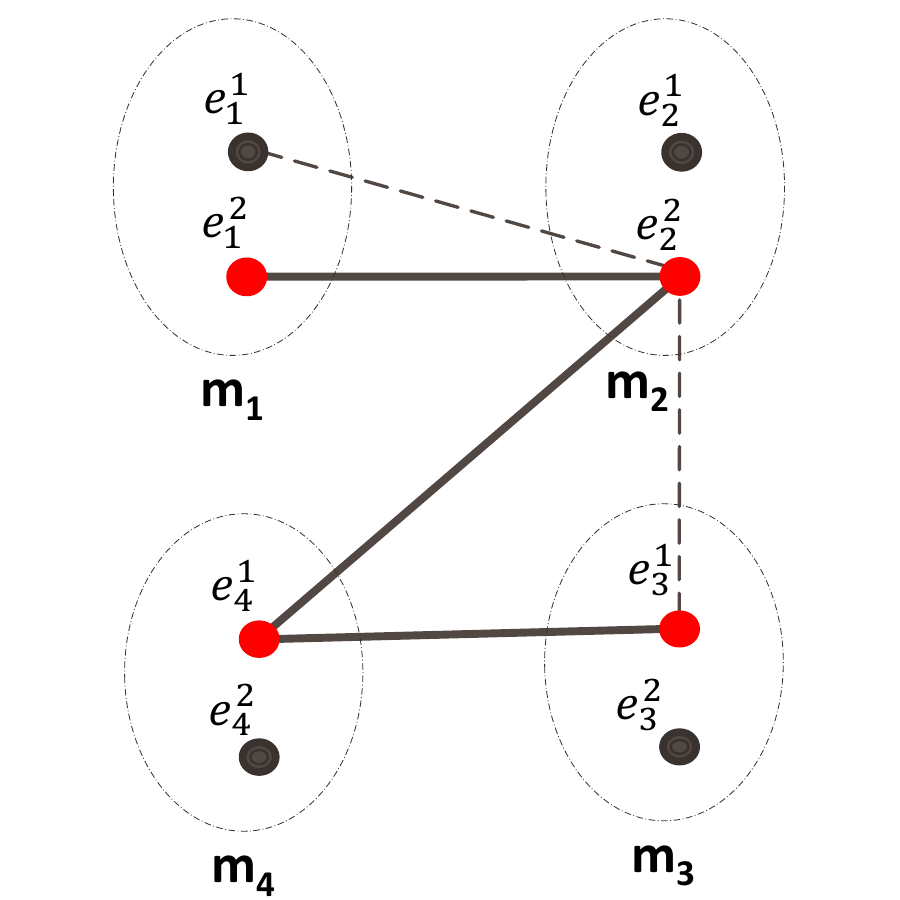}
	\caption{Entity coherence graph for a document with 4 mentions, each has 2 candidate entities. The weight of the minimum-spanning tree obtained from the selected entity set will be used to measure the coherence.}
	\label{fig:MINTREE}
\end{figure}
Given N mentions and N subsets $C_1,...,C_N\subset W$ where each $C_i$ represents the candidate entity set for mention $m_i$, an undirected \textit{entity coherence graph} $G(V,E)$ is defined as follows. The set of vertices V contains all the candidate entities in $C_1,...,C_N$. The edges in $E$ connect between two candidate entities: $e_i\in C_i$  and $e_j\in C_j$ (with $i\neq j$). The edges are weighted by the semantic distance which is computed from the local confidence and pairwise coherence scores:

\begin{equation}
d(e_i,e_j) = 1-\frac{\phi(m_i,e_i) + \psi(e_i,e_j) + \phi(m_j,e_j)}{3}
\label{eq:coher_edge_weight}
\end{equation}

The edge weights defined in this manner not only reflect the semantic relatedness distance between the two candidate entities but also encode the local confidence of a pair of assignments $m_i\mapsto e_i$ and $m_j\mapsto e_j$. We aim to find in each subset $C_i$ an entity $e_i$ such that the MINTREE coherence score of the selected entity set $\Gamma=\{e_1,...,e_N\}$ is minimized. 

The MINTREE problem defined above is equivalent to finding the minimum spanning tree on an N-partite graph G such that each of N subsets has one representative in the tree. However, for entity disambiguation task, the desired output is the selected entity set $\Gamma$, although its associated minimum spanning tree can be derived easily from $\Gamma$.

An illustration of a MINTREE output is shown in Figure~\ref{fig:MINTREE}. In this example, the document contains 4 mentions and 4 associated sets of candidate entities. The disambiguated entity for each mention is highlighted (in red) and a sample of the spanning tree is illustrated by the solid edges. The weight of the spanning tree is used to measure the coherence of the selected entity set.

\begin{table*}
	\centering
	\caption{The Spearman's rank-order correlations between the disambiguation quality (represented by the number of correct linking decisions) and three objective scores. The correlations are averaged across datasets. The results are reported with three relatedness measures: Wikipedia Link-based Measure (WLM), Normalized Jaccard Similarity (NJS) and Entity Embedding Similarity (EES).}
	\label{tb:objectives_correlation_EES}
	\begin{tabular}{l||ccc|ccc|ccc}
		\toprule
		\multicolumn{1}{c||}{\multirow{2}{*}{\textbf{Spearman's Correlation}}} & \multicolumn{3}{c|}{\textbf{WLM}} & \multicolumn{3}{c|}{\textbf{NJS}} & \multicolumn{3}{c}{\textbf{EES}} \\
		\multicolumn{1}{c||}{}                                      & ALL-L & SINGLE-L & MINTREE & ALL-L   & SINGLE-L  & MINTREE  & ALL-L & SINGLE-L & MINTREE \\
		\cmidrule(l){1-10}
		Disambiguation quality                                    & 0.924   & 0.925      & -0.927  & 0.954     & 0.952       & -0.951   & 0.947   & 0.945      & -0.947  \\ \midrule 
		ALL-Link                                                  & --       & 0.986       & -0.983   & --         & 0.995        & -0.994    & --       & 0.989       & -0.990   \\
		SINGLE-Link                                               &          & --         & -0.985   &            & --           & -0.992    &          & --          & -0.986   \\
		MINTREE                                                   &          &             & --      &            &              & --       &          &             & --      \\
		\bottomrule
	\end{tabular}
\end{table*}

Using the MINTREE coherence measure has the advantage of flexibility. It is capable of modeling complicated situations such as sparse-context documents or social texts where the documents may contain non-salient entities or entities that are not densely related in the knowledge base. In the following section, we will present a quantitative study of MINTREE and show that it is as good as other conventional models in the disambiguation task.

\paratitle{Quantitative Study of MINTREE.}
It is undoubted that the objective score of a coherence model should be correlated to the disambiguation quality. Specifically, given a set of disambiguated entities within a document, MINTREE objective score has to be lowered as the number of correct mention-entity assignments increases. We simulate the disambiguation quality by considering N+1 disambiguation results which have the number of correct assignments increasing from 0 to N:

\begin{itemize}
	\item The first disambiguation result has all mentions linking to all wrong entities. 
	\item The second disambiguation result differs with the first result by having the first mention linking to its correct entity.
	\item The $k^{th} (2<k\leq N+1)$ result differs with the ${(k-1)}^{th}$ result by having the ${(k-1)}^{th}$ mention linking to its correct entity. 
\end{itemize}	

We calculate the MINTREE objective score associated with each of the N+1 results. Spearman's correlation is calculated from the list of objective scores and the numbers of correct decisions made in N+1 disambiguations. In the ideal case, the rank-based correlation should be equal to -1 because the MINTREE score should be inversely correlated with the disambiguation quality. 
We also analyse the Spearman's correlation with ALL-Link objective (Equation~\ref{eq:all_link_objective}) as well as SINGLE-Link objective (Equation~\ref{eq:sigle_link_objective}), in the same manner. Furthermore, to show that MINTREE is correlated with other objective models, we study the correlation between each pair of the objectives.

The results are reported in Table~\ref{tb:objectives_correlation_EES}. It shows that the Spearman's correlation score between MINTREE and the disambiguation result is as high as the other objectives. The score is about 0.92 for WLM measure and more than 0.94 for NJS and EES measures. Moreover, MINTREE is highly correlated to ALL-Link and SINGLE-Link. The pairwise correlation scores are more than 0.98 across different relatedness measure. We conclude that, MINTREE is reasonably as good as other objectives when being used to model the disambiguation quality.

We also want to note that the correlations between the objective score and the disambiguation quality by WLM measure are lower than the ones by NJS and EES measures. Therefore, we will expect NJS and EES to be more effective when being used as a relatedness measure for a collective linking algorithm. We will be back to this discussion in experiment section.  Next, we will present Pair-Linking, a heuristic solution for the MINTREE problem.

%
%

\section{Pair-Linking}
\label{sec:pairlink}

\paratitle{Idea.} As mentioned earlier, finding the set of disambiguated entities $\Gamma$ is equivalent to finding the minimum spanning tree representative. Two well-known algorithms for finding minimum spanning tree  in a general graph is Kruskal's~\cite{kruskal1956shortest} and Prim's~\cite{prim1957shortest}. However, the special setting of MINTREE problem makes any direct application of Kruskal's or Prim's becoming infeasible. In this section, we introduce Pair-Linking, a heuristic to solve the MINTREE problem by finding its associated minimum spanning tree representative.

Similar to the Kruskal's algorithm, the main idea of Pair-Linking is iteratively taking an edge with the smallest distance into consideration. Specifically, Pair-Linking works on the entity coherence graph G (see the problem statement, Section~\ref{sec:model}). It iteratively takes an edge of the least possible distance that connects two entities $e_i^x$, $e_j^y$ (in two candidate sets $C_i$ and $C_j$ respectively) to form the tree. The difference compared to the original Kruskal's algorithm is that after $e_i^x$ is selected, Pair-Linking removes other vertex $e_i^{\bar{x}}$ from G such that $e_i^{\bar{x}}\neq e_i^x \wedge e_i^{\bar{x}}\in C_i$. Similar removal is done with $e_j^y$. The removing steps will ensure that there will be no other entities within the same candidate set being selected. The algorithm stops when every candidate set has one entity being selected.

Intuitively, each step of Pair-linking aims to find and resolve the most confident pair of mentions (represented by the least weighted edge on the entity coherence graph G). Furthermore, once the edge $(e_i^x, e_j^y)$ is selected, it implies that the mentions $m_i$ and $m_j$ are disambiguated to the entities $e_i^x$ and $e_j^y$ respectively.

Our Pair-Linking algorithm approximates MINTREE solution by simulating the Kruskal's but not the Prim's algorithm. The reason is twofold. First, instead of building the MST by merging smaller trees (like Kruskal's algorithm), Prim's grows the tree from a root. However, the strategy is less effective than Kruskal's in the entity disambiguation task because (Kruskal-like) Pair-linking performs disambiguation by the order of confidence score, enforcing the subsequent and less confident decisions to be consistent with the previously made and more confident assignments. This strategy has also been used in other works~\cite{ZwicklbauerSG16, RatinovRDA11, LiSD13} and been shown to improve the disambiguation performance noticeably. Another advantage of Kruskal-like over Prim-like approach is that if the coherence graph is not well-connected (sparse), the Kruskal-like Pair-Linking algorithm will return multiple coherent trees (see Figure~\ref{fig:forest_form}). Therefore, it can effectively model the sparse and noisy context.

\paratitle{Pair-Linking Example.}
\begin{figure}
	\centering
	\includegraphics[scale=0.65,,trim=2 2 2 2,clip]{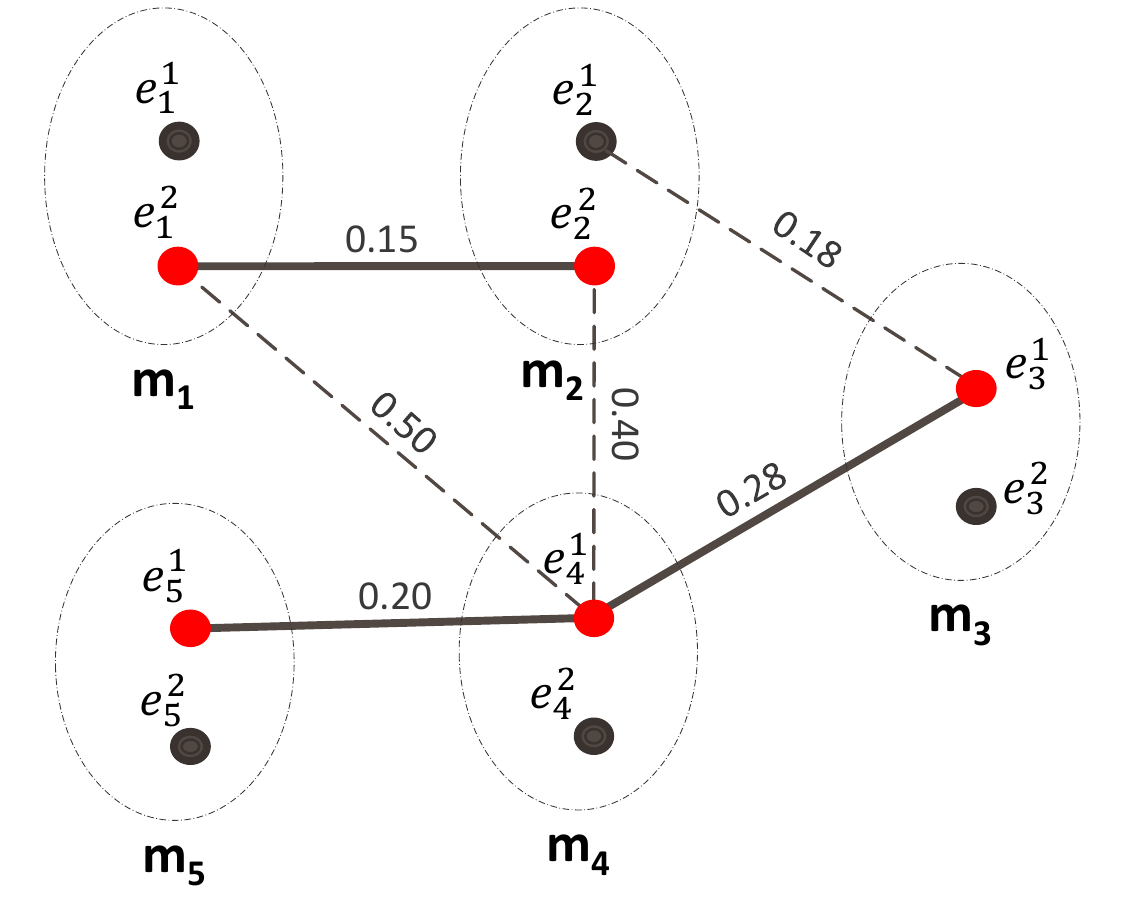}
	\caption{An example of entity coherence graph with 5 mentions, each mention has 2 candidate entities. The edges between candidate entities are weighted by the semantic distance. Only the edges with the lowest semantic distances are illustrated. The solid edges are the ones selected by the Pair-Linking process.}
	\label{fig:Pair_Linking_example}
\end{figure}
We illustrate the Pair-Linking process through an example shown in Figure~\ref{fig:Pair_Linking_example}. In this example, the given document consists of 5 mentions, each mention has 2 candidate entities. The edges between entities are weighted by the semantic distance. Pair-Linking traverses through the list of edges by the order of their weights. In the first step, Pair-Linking considers the edge with the lowest semantic distance $(e_1^2,e_2^2)$ and makes a pair of linkings with the highest confidence: $m_1 \mapsto e_1^2$ and $m_2 \mapsto e_2^2$. The edge with the second lowest semantic distance is $(e_2^1,e_3^1)$. However, since $m_2$ is already disambiguated (to $e_2^2$), any entity other than $e_2^2$ will be removed from $m_2$'s candidates, including its edges. Therefore, the next edge to be considered is $(e_4^1,e_5^1)$. As a result, $m_4$ and $m_5$ are disambiguated to $e_4^1$ and $e_5^1$ respectively. Lastly, $(e_3^1,e_4^1)$ is taken into account and one additional linking is made \ie $m_3 \mapsto e_3^1$. Pair-Linking stops at this step because all the 5 mentions are already disambiguated to its associated entities (highlighted in red in Figure~\ref{fig:Pair_Linking_example}). Note that for entity disambiguation task, it is not necessary to derive the minimum spanning tree associated with the set of selected entities, although it can be done by continuing picking up latter edges until a connected tree is formed. 

\begin{algorithm}[t]
	\caption{Pair-Linking algorithm}
	\label{alg:pairlink}
	\DontPrintSemicolon
	\SetKwInOut{Input}{input}
	\SetKwInOut{Output}{output}
	\Input{$N$ mentions $(m_1,...,m_N)$. Mention $m_i$ has candidate set $\{C_i \subset W \}$}
	\Output{$\Gamma = (e_1,...,e_N)$}
	
	\SetKwFunction{Fps}{pair\_score}
	\SetKwFunction{mcp}{most\_confident\_pair}
	\SetKwFunction{tpf}{top\_pair}
	\SetKwProg{Fn}{Function}{:}{}
	
	$e_i \leftarrow null,$  $\forall e_i \in \Gamma$
	
	\For{each pair $(m_i,m_j) \wedge m_i\ne m_j$}{
		$Q_{m_i,m_j} \leftarrow \tpf(m_i,C_i,m_j,C_j)$
		
		$Q$.add( $Q_{m_i,m_j}$ )
	}
	
	\While{$(\exists e_i \in \Gamma, e_i = null)$}{
		$(m_i,e_i^x,m_j,e_j^y) \leftarrow \mcp(Q)$
		
		$e_i^x \mapsto e_i$ (Disambiguate $m_i$ to $e_i^x$)
		
		$e_j^y \mapsto e_j$ (Disambiguate $m_j$ to $e_i^x$)
		
		\For{$k :=1 \to N \wedge e_k = null$}{						
			$Q_{m_k,m_i} \leftarrow \tpf(m_k,C_k,m_i,\{e_i\})$
			
			$Q_{m_k,m_j} \leftarrow \tpf(m_k,C_k,m_j,\{e_j\})$
		}
	}
\end{algorithm}

\paratitle{Pair-Linking Procedure.} 
We detail Pair-Linking procedure in Algorithm~\ref{alg:pairlink}. Specifically, Pair-Linking maintains a priority queue $Q$ and each element $Q_{m_i,m_j}$ tracks the most confident linking pairs involving mentions $m_i$ and $m_j$. $Q_{m_i,m_j}$ is initialized by calling  function $top\_pair(m_i,C_i,m_j,C_j)$, where $C_i$ is the set of candidate entities that mention $m_i$ can link to. The function returns a pair assignment $m_i\mapsto e_i^x$ and $m_j\mapsto e_j^y$, such that $e_i^x \in C_i$, $e_j^y \in C_j$, and the confidence score of the pair assignment is the highest among $C_i \times C_j$ (\ie the edge distance is the smallest according to Equation~\ref{eq:coher_edge_weight}). After initialization, Pair-Linking iteratively retrieves the most confident pair assignment from $Q$ (Line 7) and links the pair of mentions to the associated entities (Lines 8-9). After that, Pair-Linking updates $Q$,  more precisely, $Q_{m_k,m_i}$ and $Q_{m_k,m_j}$ (Lines 10-13). For $Q_{m_k,m_i}$, the possible pairs of assignments between $m_k$ and $m_i$ are now conditioned by $m_i\mapsto e_i^x$, and the same applies to $Q_{m_k,m_j}$.

\paratitle{Early Stop.}
The most expensive part of the algorithm is the initialization of $Q$ which requires computing $top\_pair$ between every two mentions. A straightforward implementation of the function $top\_pair(m_i,C_i,m_j,C_j)$ will scan through all possible candidate pairs between the two mentions. It has the time complexity of $\mathcal{O} \left( k^2 \right)$ where $k$ is the number of candidates per mention. This leads to an overall complexity of $\mathcal{O} \left(N^2k^2\right)$ for the Q's initialization (Lines 2-5). Here, $N$ is the number of mentions. However, since only the pair of candidates with the highest confidence score is recorded for a pair of mentions $m_i$ and $m_j$, Pair-Linking uses \textit{early stop} to avoid scanning through all  possible candidate pairs. Specifically, it sorts each of N candidate set by the local scores ($\mathcal{O} \left(Nk\log{}k\right)$) and traverses the sorted list in descending order. Early stop is applied if the current score is worse than the highest score by a specific margin, \ie the largest possible value of $\psi(e_i,e_j)$, see Equation~\ref{eq:coher_edge_weight}. 

In the best case, if early stop is applied right after getting the first score, the complexity of $top\_pair(m_i,C_i,m_j,C_j)$ is  $\mathcal{O} \left( 1 \right)$ and the overall time complexity becomes $\mathcal{O} \left(N^2+Nk\log{}k\right)$. Indeed, early stop significantly reduces the running time of Pair-Linking in practice while still maintaining the correctness of the algorithm.

\section{Experiment}
\label{sec:exp}

We use  Wikipedia dump on 01-Jul-2016 as the target knowledge base. It consists of  5,187,458 entities. In the following subsections, we will describe the experiment setting protocol, datasets, and methods in comparison. Lastly, we present and discuss the experiment results.

\subsection{Experimental Setting}
\label{ssec:setup}

\paratitle{Candidate Generation and Filtering.}
As a common approach~\cite{ShenWH15,ZwicklbauerSG16, PappuBMST17}, our candidate generation is purely based on the textual similarity between a mention's surface form and an entity's title including all its variants. We use a dictionary based technique for candidate retrieval~\cite{ShenWH15}. The dictionary is built by exploiting entity titles, anchor texts, redirect pages, and disambiguation pages in Wikipedia. If a given mention does not present in the dictionary, we use its n-grams to retrieve the candidates. We further improve the recall of candidate generation by correcting the mention's boundary. In several situations, a given mention may contain trivial words (\eg the, Mr., CEO, president) that are not indexed by the dictionary. We use an off-the-shelf Named Entity Recognizer (NER)\footnote{\scriptsize We used the Standford NER tool in this work.} to refine the mention's boundary in these cases. As in~\cite{GottipatiJ11}, we also utilize the NER output to expand the mention's surface form. Specifically, if mention $m_1$ appears before $m_2$ and $m_1$ contains $m_2$ as a substring, we consider $m_1$ as an expanded form of $m_2$, and candidates of $m_1$ will be included into the candidate set of $m_2$.

We train a Gradient Boosted Regression Trees model~\cite{friedman2001greedy} as the candidate ranker used to reduce the size of the candidate set. For each pair of mention and candidate entity, \ie $(m,e)$, we use the following statistical and lexical features for ranking.

\begin{itemize}
	\item Prior probability $P(e|m)$.   $P(e|m)$ is the likelihood that a mention m $m$ will be mapped to an entity $e$. $P(e|m)$ is pre-calculated based on the hyperlinks in Wikipedia.
	\item String similarity. We use several string similarity measures: (i) edit distance, (ii) whether mention $m$ exactly matches entity $e$'s name, (iii) whether $m$ is a prefix or suffix of the entity name, and (iv) whether $m$ is an abbreviation of the entity name. Note that the string similarity features are calculated for the original mention, boundary-corrected mention, and expanded mention.
\end{itemize}

We use IITB labeled dataset~\cite{KulkarniSRC09} to train the ranking model. For each mention, we take the top 20 scored entities as its candidate set. Taking fewer candidate entities will lead to low recall while using more candidates will degrade disambiguation accuracy in the later step. Similar observations are also reported in~\cite{ZwicklbauerSG16, GaneaGLEH16}.

Note that the candidate ranker described above is different from the model used to estimate the local confidence score presented in Section~\ref{sec:preliminaries}. The former aims to maximize the recall of top-k ranked candidates while the latter targets on the accuracy of prediction, \ie the top-1 ranked candidate. 

\paratitle {Local Confidence Score and Pairwise Coherence Score.} We use the local score which is  the output of a learning to rank model (see Section~\ref{sec:preliminaries}). Furthermore, for pairwise coherence, we study and report the results with three kinds of measures: Wikipedia link-based measure (WLM), normalized Jaccard similarity (SNS) and entity embedding similarity (EES). In addition, we use a hyper-parameter $\beta$ to control the contribution between the local confidence and the pairwise coherence components in the final objective. For example, the refined objective for Equation~\ref{eq:all_link_objective} can be written as follows:

\begin{equation}
\small
\Gamma^* = \argmax_{\Gamma} \left[(1-\beta)\sum_{i=1}^N \phi(m_i,e_i) + \beta\sum_{i=1}^N \sum_{j=1, j\neq i}^N \psi(e_i,e_j)\right]\label{eq:common_objective_with_beta}
\end{equation}

\paratitle {Cross Validation.}
We use 5-fold cross validation in evaluation. At each iteration of the cross validation, the learning to rank model GBT and the parameter $\beta$  are learned based on 4 training partitions. The best setting is then used to perform disambiguation on the remaining test partition. The final disambiguation result is the aggregation of all predictions in 5 iterations of the cross validation.

\subsection{Datasets and Methods in Comparison}
\label{ssec:datasets_methods}
\begin{table}
	\centering
	\caption{Statistics of the 8 test datasets used in our evaluation. ${|D|}$, $|M|$, $Avg_m$, and $Length$ are number of documents, number of mentions, average number of mentions per document, and average number of words per document, respectively.}
	\label{tb:datasets}
	\begin{tabular}{llrrrr}
		\toprule
		\textbf{Dataset}              & \textbf{Type} & \textbf{$|D|$} & \textbf{$|M|$} & \textbf{$Avg_m$} & \textbf{Length} \\
		
		\midrule
		Reuters128	&news	& 111	& 637	& 5.74	& 136\\
		ACE2004		&news	& 35 	& 257	& 7.34	& 375\\
		MSNBC		&news	& 20 	& 658	& 32.90	& 544\\
		DBpedia 	&news	& 57 	& 331	& 5.81	& 29\\
		RSS500		&RSS-feeds	& 343 	& 518	& 1.51	& 30\\
		KORE50		&short sentences	& 50 	& 144	& 2.88	& 12\\
		Micro14	&tweets	& 696 	& 1457	& 2.09	& 18\\
		AQUAINT		&news	& 50	&726	&14.52	&220\\
		\bottomrule
	\end{tabular}
\end{table}

\paratitle{Datasets.} We evaluate the performance on 8 benchmark datasets coming from different domains, including short and long text, formal and informal text. The statistics of each dataset is represented in Table~\ref{tb:datasets}. Note that, we only consider the mentions whose linked entities appear in Wikipedia; the same setting has been used in~\cite{PappuBMST17,GaneaGLEH16,yamada2016joint,ZwicklbauerSG16}. We describe each dataset as follows:

\begin{itemize}
	\item \textbf{Reuters128}~\cite{roder2014n3} contains 128 economic news articles taken from the Reuters-21587 corpus. There are 111 documents containing linkable mentions (based on Wikipedia 01-Jul-2016 dump).
	
	\item \textbf{ACE2004}~\cite{RatinovRDA11}  is a subset of ACE2004 co-reference documents annotated by Amazon Mechanical Turk. It has 35 documents, each has 7 mentions on average.
	
	\item \textbf{MSNBC}~\cite{Cucerzan07} is created from MSNBC news articles. It contains 20 documents, each has 33 mentions on average. The dataset includes many entities that can be linked via direct relation in DBpedia. Therefore, many disambiguation systems can easily achieve high accuracy on this dataset.
	
	\item \textbf{DBpedia Spotlight (DBpedia)} is a news corpus and contains many non-named entity mentions such as parents, car, dance. It is an average-size dataset in which each document contains 5 to 6 mentions on average.
	
	\item \textbf{RSS500}~\cite{GerberHBSUN13} is RSS feeds - a short formal text collection covers a wide range of topics \eg world, business, science, etc. The dataset is one of N3 datasets~\cite{roder2014n3} which are carefully created as a benchmark to evaluate named entity disambiguation systems.
	
	\item \textbf{KORE50}~\cite{HoffartSNTW12} contains 50 short sentences on various topics \eg music, celebrities, and business. Most mentions are the first names referring to persons and they are highly ambiguous. It is considered as a challenging dataset.
	
	\item \textbf{Microposts2014 (Micro14)}~\cite{BasaveRVRSD14} is a collection of tweets,  introduced in the `Making Sense of Microposts 2014' challenge. The textual context for a document is very limited and noisy due to the nature of tweet.
	The dataset has train/test partitions. We use the test partition in the evaluation so that results can be compared to other works.
	
	\item \textbf{AQUAINT}~\cite{MilneW08} contains 50 news documents from Xinhua News Service, the New York Times and Associated Press news corpus.
\end{itemize}

\paratitle{Collective Linking Methods.} We compare our Pair-Linking algorithm with the following state-of-the-art collective linking (CL) algorithms.

\begin{itemize}
	\item \textbf{Iterative Substitution} (\textbf{Itr\_Sub (AL)})~\cite{ShenWLW12} is an approximate solution for the ALL-Link objective (Equation~\ref{eq:all_link_objective}). Each mention is initially assigned to a candidate entity which has the highest local score. The algorithm iteratively substitutes an assignment $m_i \mapsto e_i^x$ with another mapping $m_i \mapsto e_j^y$ as long as it improves the objective score. We also study the performance of Iterative Substitution with the Sing-Link objective (Equation~\ref{eq:sigle_link_objective}) and refer to it as \textbf{IterSub (SL)}.
	
	\item \textbf{Loopy Belief Propagation} (\textbf{LBP(AL)})~\cite{GaneaGLEH16,GlobersonLCSRP16} solves the inference problem (Equation~\ref{eq:all_link_objective}) through loopy belief propagation technique~\cite{MurphyWJ99}. Similar to the Iterative Substitution algorithm, we also study another setting with the SINGLE-Link objective and refer to it as \textbf{LBP(SL)}.
	
	\item \textbf{Forward-Backward} (\textbf{FwBw})~\cite{austin1991forward}  considers only the local coherence in the disambiguation objective. It uses dynamic programming to derive the optimal assignments. The work in~\cite{PappuBMST17} shows that the approach is effective and efficient for entity extraction in queries.
	
	\item \textbf{Densest Subgraph} (\textbf{DensSub})~\cite{HoffartYBFPSTTW11} applies dense subgraph algorithm to prune the mention-candidate graph. Subsequently, local search is performed to derive the mention-entity assignment based on an objective function similar to ALL-Link. 
	
	\item \textbf{Personalized PageRank} (\textbf{PageRank}) is used by DoSeR~\cite{ZwicklbauerSG16}. It performs personalized PageRank on a mention-candidate graph and uses the stabilized scores for disambiguation. Additionally, DoSeR introduces a 'pseudo` topic node to enforce the coherence between disambiguated entities and the main topic's context.
\end{itemize}

We acknowledge a relevant work in \cite{GlobersonLCSRP16} also addresses the issue of mentioned entities that are  not salient or not well-connected in KB. To perform collective linking, the authors propose a model that considers only top-k most related connections for each entity. However, the model is trained in end-to-end fashion together with the parameters for local confidence and coherence scores. In contrast, our work only focuses on the collective linking component and uses existing local similarity and pairwise coherence measures. Therefore a comparison to their work is not included in our study.

\paratitle{Evaluation Measures.}
To evaluate the performance of different collective linking methods, we install Gerbil benchmarking framework~\cite{UsbeckRNBBBCCCE15} (Version 1.2.4)  and run the evaluation locally. We report the disambiguation results by the widely used measures: Precision, Recall, and F1. Specifically, let $\Gamma_g$ be the set of ground-truth assignments and $\Gamma^*$ be the mappings produced by a disambiguation system, the evaluation metrics are expressed as follows: 

$P = \frac{|\Gamma^* \cap \Gamma_g|}{|\Gamma^*|}\hspace{0.4in}R = \frac{|\Gamma^* \cap \Gamma_g|}{|\Gamma_g|}\hspace{0.4in}F1 = \frac{2\times PR}{P+R}$


%



\begin{table*}
	\centering
	\caption{Micro-averaged F1 of different collective linking algorithms with different coherence measures. The best scores are in boldface and the second-best are underlined. The number of win and runner-up each method performs across different datasets are also illustrated. Significance test is performed on Reuters123, RSS500 and Micro14 datasets (denoted by $^*$) which contain a sufficient number of documents. $\dagger$~indicates the difference against the Pair-Linking's F1 score is statistically significant by one-tailed paired $t$-test (with $p < 0.05$).}
	\label{tb:CL_perfomrnace}
	
	\subfloat [WLM as coherence measure.
	]{
		\begin{tabular}{l|cccccccc|ccr}
			\toprule
			CL Method     & Reuters128$^*$     & ACE2004        & MSNBC          & Dbpedia        & RSS500$^*$         & KORE50         & Micro14$^*$ & AQUAINT        & Average        & \#1st & \#2nd\\
			\midrule
			Iter\_Sub(AL) & 0.795          & {\ul 0.873}    & 0.809          & 0.821          & 0.775$^\dagger$          & 0.506          & 0.798          & 0.857         & 0.779          &0&1\\
			Iter\_Sub(SL) & 0.778$^\dagger$          & 0.849          & \textbf{0.874} & 0.827          & 0.758$^\dagger$          & 0.484          & 0.794          & 0.849         & 0.777          &1&0\\
			LBP(AL)       & {\ul 0.800}    & 0.867          & 0.847          & 0.837          & {\ul 0.776}    & 0.487          & 0.798          & 0.855          & 0.783          &0&2\\
			LBP(SL)       & 0.793          & 0.865          & 0.850          & 0.828          & 0.772          & 0.496          & \textbf{0.805} & \textbf{0.868} & 0.785          &2&0\\
			FwBw          & 0.788          & \textbf{0.876} & 0.850          & \textbf{0.844} & 0.772$^\dagger$          & {\ul 0.526}    & {\ul 0.799}    & 0.859          & {\ul 0.789}    &2&2\\
			DensSub     &0.788& {\ul 0.873}&0.831&0.823&0.766$^\dagger$&0.523&0.790&0.853&0.781& 0 & 1\\
			PageRank      & 0.767$^\dagger$          & 0.832          & 0.791          & 0.722          & 0.769$^\dagger$          & 0.490          & 0.772$^\dagger$          & 0.812          & 0.744          &0&0\\
			Pair-Linking  & \textbf{0.802} & 0.871          & {\ul 0.864}    & {\ul 0.842}    & \textbf{0.785} & \textbf{0.535} & 0.796          & {\ul 0.862}    & \textbf{0.795}&3&3\\
			\bottomrule
		\end{tabular}
	}
	
	\subfloat [NJS as coherence measure.
	]{
		\begin{tabular}{l|cccccccc|ccr}
			\toprule
			CL Method     & Reuters128$^*$     & ACE2004        & MSNBC          & Dbpedia        & RSS500$^*$         & KORE50         & Micro14$^*$ & AQUAINT        & Average        & \#1st & \#2nd\\
			\midrule
			Iter\_Sub(AL) & {\ul 0.840}    & 0.877          & 0.882          & 0.810          & 0.783$^\dagger$          & 0.689          & 0.814          & 0.869          & 0.821          &0&1\\
			Iter\_Sub(SL) & 0.821          & 0.876          & 0.878          & 0.812          & {\ul 0.795}    & 0.671          & 0.812          & 0.859          & 0.815          &0&0\\
			LBP(AL)       & 0.839          & 0.883          & 0.883          & 0.825          & 0.790          & {0.728}    & 0.812          & {\ul 0.871}    & {\ul 0.829}    &0&1\\
			LBP(SL)       & 0.813          & {\ul 0.886}    & {\ul 0.886}    & {\ul 0.833}    & 0.788          & 0.726          & \textbf{0.818} & 0.868          & 0.827          &1&3\\
			FwBw          & 0.813$^\dagger$          & 0.883          & 0.870          & \textbf{0.849} & 0.792          & {0.728}    & {\ul 0.815}    & 0.869          & 0.827          &1&1\\
			DensSub &0.835&0.881&0.855&0.820&0.778$^\dagger$&{\ul 0.731}&0.806$^\dagger$&0.853&0.820 & 0 & 1\\
			PageRank      & 0.835          & \textbf{0.897} & 0.864          & {\ul 0.833}    & 0.783          & 0.707          & 0.808          & \textbf{0.875} & 0.825          &2&1\\
			Pair-Linking  & \textbf{0.846} & 0.876          & \textbf{0.892} & 0.831          & \textbf{0.797} & \textbf{0.764} & 0.814          & 0.870          & \textbf{0.836}&4&0\\
			\bottomrule
		\end{tabular}
	}
	
	\subfloat [Entity Embedding Similarity (EES) as coherence measure.
	]{
		\begin{tabular}{l|cccccccc|ccr}
			\toprule
			CL Method     & Reuters128$^*$     & ACE2004        & MSNBC          & Dbpedia        & RSS500$^*$         & KORE50         & Micro14$^*$ & AQUAINT        & Average        & \#1st & \#2nd\\
			\midrule
			Iter\_Sub(AL) & {\ul 0.852}    & \textbf{0.905} & 0.875          & 0.837          & 0.795          & 0.556          & 0.806          & 0.872          & 0.812          &1&1\\
			Iter\_Sub(SL) & 0.807$^\dagger$          & 0.871          & 0.864          & 0.820          & 0.801          & 0.565          & {\ul 0.809}    & 0.860          & 0.800          &0&1\\
			LBP(AL)       & {\ul 0.852}    & 0.884          & \textbf{0.897} & \textbf{0.851} & 0.801          & 0.581          & {\ul 0.809}    & {\ul 0.877}    & 0.819          &2&3\\
			LBP(SL)       & 0.846          & {\ul 0.889}    & 0.882          & 0.836          & 0.802          & {\ul 0.631}    & \textbf{0.817} & 0.872          & {\ul 0.822}    &1&2\\
			FwBw          & 0.834$^\dagger$          & 0.885          & 0.891          & {\ul 0.850}    & {\ul 0.805}    & 0.587          & {\ul 0.809$^\dagger$}    & 0.870          & 0.816          &0&3\\
			DensSub      &0.825$^\dagger$&0.836&0.840&0.805&0.796$^\dagger$&0.586&0.779$^\dagger$&0.858&0.791 & 0 & 0\\
			PageRank      & 0.817$^\dagger$          & 0.874          & 0.877          & 0.827          & 0.768$^\dagger$          & 0.503          & 0.790$^\dagger$          & 0.860          & 0.789          &0&0\\
			Pair-Linking  & \textbf{0.856} & 0.879          & {\ul 0.894}    & 0.846          & \textbf{0.806} & \textbf{0.637} & \textbf{0.817} & \textbf{0.885} & \textbf{0.827}&5&1\\
			\bottomrule
		\end{tabular}
	}
	
	\subfloat [Combination of NJS\&EES as coherence measure.
	]{
		\begin{tabular}{l|cccccccc|ccr}
			\toprule
			CL Method     & Reuters128$^*$     & ACE2004        & MSNBC          & Dbpedia        & RSS500$^*$         & KORE50         & Micro14$^*$ & AQUAINT        & Average        & \#1st & \#2nd\\
			\midrule
			Iter\_Sub(AL) & 0.856          & {\ul 0.894}    & 0.879          & 0.839          & 0.793$^\dagger$          & 0.682          & 0.811          & 0.876          & 0.829           &0&1\\
			Iter\_Sub(SL) & 0.807$^\dagger$          & 0.883          & 0.870          & 0.835          & 0.809          & 0.653          & 0.808          & 0.850          & 0.814           &0&0\\
			LBP(AL)       & \textbf{0.864} & 0.861          & 0.895          & 0.833          & 0.777$^\dagger$          & 0.715          & {\ul 0.822}    & {0.877}    & 0.831           &1&1\\
			LBP(SL)       & 0.823$^\dagger$          & 0.875          & 0.900    & {\ul 0.843}    & {\ul 0.814}    & {\ul 0.762}    & \textbf{0.824} & 0.872          & {\ul 0.839}     &1&3\\
			FwBw          & 0.830$^\dagger$          & \textbf{0.895} & {\ul 0.905} & 0.832          & 0.802$^\dagger$          & 0.749          & 0.818          & 0.866          & 0.837           &1&1\\
			DensSub      &0.851&0.886&0.887&0.835&0.806$^\dagger$&0.738&0.809&{\ul 0.878}& 0.836 & 0 & 1\\
			PageRank      & 0.837$^\dagger$          & 0.882          & 0.888          & 0.822          & 0.785$^\dagger$          & 0.512          & 0.797$^\dagger$          & 0.872          & 0.799           &0&0\\
			Pair-Linking  & {\ul 0.859}    & 0.883          & \textbf{0.910}          & \textbf{0.845} & \textbf{0.823} & \textbf{0.787} & 0.813          & \textbf{0.879} & \textbf{0.850} &5&1\\
			\bottomrule
		\end{tabular}
	}
	
\end{table*}

For all the measures, we report the micro-averaged score (\ie aggregated across mentions not documents), and use the micro-averaged $F1$ as the main metric for comparison.

\subsection{Result and Discussion}
\label{ssec:result_discussion}

\subsubsection{Collective linking performance.}

We study the performance of different collective linking algorithms with different settings of coherence measures. The result is listed in Table~\ref{tb:CL_perfomrnace}. Note that in this experiment, we use a Gradient Boosting model to estimate the local confidence score (see Section~\ref{sec:preliminaries}). This is different from our previous work~\cite{minh2017neupl} where we utilize a deep neural network model for the estimation. Therefore the result in this table is slightly different from the former one.

As illustrated  in Table~\ref{tb:CL_perfomrnace}, the coherence measure significantly affects the performance of all collective linking algorithms. The Normalized Jaccard Similarity (NJS) and entity embedding similarity (EES) are shown to be more effective than the Wikipedia Link-based Measure (WLM). Furthermore, we try to combine different measures by taking their average coherence scores. Overall, the combination involved two former measures (\ie NJD and EES) works the best. The combined scheme outperforms other individual scheme as shown in Table~\ref{tb:CL_perfomrnace}.

The approximation algorithm Loopy Belief Propagation (LBP) is consistently better than the Iterative Substitution in both objective settings ALL-Link (AL) and SINGLE-Link (SL). Furthermore, comparing between ALL-Link and SINGLE-Link, Iterative Substitution and LBP algorithms give comparable performance across different coherence measures.

\begin{table*}
	\centering
	\caption{Average time to disambiguate mentions in one document (in milliseconds) for each dataset. The time for preprocessing steps such as candidate generation is not included. The numbers of win and runner-up are also illustrated.}
	\label{tb:CL_running_time}
	\begin{tabular}{l|rrrrrrrr||rr}
		\toprule
		\textbf{CL method} & \textbf{Reuters128} & \textbf{ACE2004} & \textbf{MSNBC} & \textbf{Dbpedia} & \textbf{RSS500} & \textbf{KORE50} & \textbf{Micro14} & \textbf{AQUAINT} & \#1st & \#2nd\\
		\midrule
		Iter\_Sub(AL) & 97.515     & 21.369  & 3010.214 & 12.922  & 0.127  & 2.235  & 0.682     & 293.271 & 0 & 0\\
		Iter\_Sub(SL) & 67.772     & 20.183  & 3211.341 & 11.603  & 0.108  & 2.284  & 0.684     & 107.640 & 0 & 0\\
		LBP(AL)       & 40.049     & 41.911  & 1584.504 & 42.673  & 0.331  & 11.515 & 3.667     & 269.854 & 0 & 0\\
		LBP(SL)       & 92.625     & 43.173  & 4421.172 & 44.263  & 0.289  & 8.627  & 3.170     & 403.140 & 0 & 0\\
		FwBw          & \textbf{0.940}      & {\ul 1.975}   & \textbf{8.880}    & {\ul 2.034}   & {\ul 0.103}  & {\ul 1.190}  & {\ul 0.367}     & {\ul 4.959}  & 2 & 6\\		
		DensSub      &166.862&221.437&12714.782&168.716&1.196&13.719&7.402&1121.231& 0 & 0\\
		PageRank      & 110.572    & 77.398  & 4293.670 & 132.009 & 5.436  & 64.982 & 15.796    & 375.239 & 0 & 0\\
		Pair-Linking  & {\ul 1.721}      & \textbf{0.590}   & {\ul 28.699}   & \textbf{0.491}   & \textbf{0.025}  & \textbf{0.951}  & \textbf{0.117}     & \textbf{3.105} & 6 & 2\\
		\bottomrule
	\end{tabular}
\end{table*}

\begin{table*}
	\centering
	\caption{Micro-averaged $F1$ of Pair-Linking (using NJS\&EES coherence measure) and other disambiguation systems. The best results are in boldface and the second-best are underlined.}
	\label{tb:PL_with_baselines}
	\begin{tabular}{l|cccccccc|r}
		\toprule		
		\textbf{System}                 & \textbf{Reuters128} & \textbf{ACE2004} & \textbf{MSNBC} & \textbf{Dbpedia} & \textbf{RSS500} & \textbf{KORE50} & \textbf{Micro14} & \textbf{AQUAINT} & \textbf{Average} \\
		\midrule
		PBoH~\cite{GaneaGLEH16}                   & 0.759          & 0.876          & 0.897    & 0.791          & 0.711          & {\ul 0.646}    & 0.725          & 0.841          & 0.781          \\
		DoSeR~\cite{ZwicklbauerSG16}                  & \textbf{0.873} & \textbf{0.921} & \textbf{0.912} & 0.816          & {\ul 0.762}    & 0.550          & 0.756          & 0.847          & {\ul 0.805}    \\
		$P(e|m)$ (local)                 & 0.697          & 0.861          & 0.781          & 0.752          & 0.702          & 0.354          & 0.650          & 0.835          & 0.704          \\
		Xgb (local)            & 0.776          & 0.872          & 0.834          & {\ul 0.818}    & 0.756          & 0.496          & {\ul 0.789}    & {\ul 0.855}    & 0.775          \\
		Pair-Linking (NJS\&EES) & 0.859          & {\ul 0.883}    & {\ul 0.910}          & \textbf{0.845} & \textbf{0.823} & \textbf{0.787} & \textbf{0.813} & \textbf{0.879} & \textbf{0.850}\\
		
		\bottomrule
	\end{tabular}
\end{table*}

Graph based algorithms such as DensSub and PageRank are sensitive to the choice of coherence measure. For example, PageRank only produces good results when working with the NJS coherence measure, \ie 0.825 F1 score versus 0.744 and 0.789 when working with WLM and EES measure, respectively. On the other hand, Pair-Linking is quite robust to all three measures. It outperforms other methods on more challenging and short text datasets such as Reuters128, RSS500, and KORE50.

Forward-Backward algorithm (FwBw) is shown to perform better on short text datasets (RSS and Micro14) in comparison to long text datasets (Reuters and AQUAINT). This is because for long documents, useful evidence may not be presented in the local context.  

\subsubsection{Collective linking running time.}

\begin{table}
	\centering
	\caption{Time complexity of different linking algorithms. $N$ is number of mentions, $k$ is average number of candidates per mention, and $\mathcal{I}$ is number of iterations for convergence.}
	\label{tb:cl_complexity}
	\begin{tabular}{l|cr}
		\toprule
		\textbf{Collective Linking}&  \textbf{Best case}& \textbf{Worst case}\\
		\midrule
		ItrSub&$\scriptstyle{\mathcal{O} \left(N^3k\right)}$				&$\scriptstyle{\mathcal{O} \left(\mathcal{I}\times N^3k\right)}$\\
		LBP&$\scriptstyle{\mathcal{O} \left(N^2k^2\right)}$			&$\scriptstyle{\mathcal{O} \left(\mathcal{I}\times N^2k^2\right)}$\\
		FwBw&$\scriptstyle{\mathcal{O} \left(Nk^2\right)}$				&$\scriptstyle{\mathcal{O} \left(Nk^2\right)}$\\
		DensSub&$\scriptstyle{\mathcal{O} \left(N^3k^2+N^2k^2\right)}$			&$\scriptstyle{\mathcal{O} \left(N^3k^2+\mathcal{I}\times N^2k^2\right)}$\\			
		PageRank&$\scriptstyle{\mathcal{O} \left(N^2k^2\right)}$		&$\scriptstyle{\mathcal{O} \left(\mathcal{I}\times N^2k^2\right)}$\\		
		Pair-Linking&$\scriptstyle{\mathcal{O} \left(Nk\log{k}+N^2\right)}$ 	&$\scriptstyle{\mathcal{O} \left(Nk\log{k}+N^2k^2\right)}$\\
		\bottomrule
	\end{tabular}
\end{table}

The theoretical time complexities of different collective linking methods are listed in Table~\ref{tb:cl_complexity}. FwBw has the lowest time complexity in worst case since it only considers  adjacent mentions. By using dynamic programming~\cite{austin1991forward}, FwBw calculates the score of each assignment $m_i \mapsto e_i$ by considering all possible states in the previous decision (\ie $m_{i-1} \mapsto e_{i-1}$), resulting in the complexity of $\mathcal{O} \left(k\right)$ where $k$ is the number of candidate entities per mention. The overall time complexity of FwBw is $\mathcal{O} \left(Nk^2\right)$ where $N$ is the number of mentions.

Not surprisingly, optimization-based (Itr\_Sub, LBP) and graph-based methods (DensSub, PageRank) have highest time complexity. While Itr\_Sub and LBP require multiple iterations to solve the optimization problems, two graph-based algorithms DensSub and PageRank work on a mostly complete entity graph with $N^2k^2$ edges. DensSub additionally requires $\mathcal{O} \left(N^3k^2\right)$ to pre-process the graph (\ie filter noisy entities by shortest path distances). Furthermore, PageRank iteratively operates on the mention-entity matrix for convergence and it leads to the complexity of ${\mathcal{O} \left(\mathcal{I}\times N^2k^2\right)}$ where $\mathcal{I}$ is the number of iterations required. On the other hand, Pair-Linking only needs to traverse all possible pairs of linking assignment (\ie $(m_{i},e_{i}), (m_{j},e_{j})$)  at most once, resulting in the complexity of $\mathcal{O} \left(N^2k^2\right)$.  Furthermore, the worst case of Pair-Linking is the prerequisite of any graph-based algorithm (\eg DensSub, PageRank) because building the mention-entity graph for $N$ mentions, each has $k$ candidate entities will require $Nk$  vertices and $N^2k^2$ edges.

It is also worth mentioning that Pair-Linking is interested in only the pairs of linking assignments having the highest confidence scores. Therefore, by using a priority queue to keep track of the top confident pairs, it can avoid traversing through every pair at each step. Empirical results show that Pair-Linking is indeed fast, partially due to ``early stop'' in implementation described in Section~\ref{sec:pairlink}.  Since only a few pairs of assignments dominate the Pair-Linking scores, a large number of pairs are ignored by the early stop. Table~\ref{tb:CL_running_time} shows that the running time of Pair-Linking (including the time used to construct the priority queue is even smaller than FwBw on 6 out of 8 datasets, making Pair-Linking the most effective and efficient linking algorithm.

Considering the long text dataset MSNBC, Pair-Linking is nearly 50-100 times faster than the next effective algorithm LBP(AL), as shown in Table~\ref{tb:CL_running_time}. FwBw is faster than Pair-Linking but its linking accuracy is worse than Pair-Linking in several datasets (see Table~\ref{tb:CL_perfomrnace}). Different from Pair-Linking, FwBw only considers local coherence in its objective function and ignores connections between entities that are far away (\eg across paragraphs). On a side note, the good results of FwBw and Pair-Linking hint that a hybrid algorithm that incorporates both FwBw and Pair-Linking's ideas can further improve the performance.

\subsubsection{Comparison with other disambiguation systems.}
\begin{table}
	\centering
	\caption{Micro-averaged Precision, Recall, $F1$ score of Pair-Linking with the combined NJS\&EES as coherence measure.}
	\label{tb:prf1_bestab}
	\begin{tabular}{l|ccc}
		\toprule
		\textbf{Data set}	&\textbf{Precision}	&\textbf{Recall}	&\textbf{F1}\\
		\midrule
		Reuters128       & 0.866              & 0.853           & 0.859       \\
		ACE2004          & 0.888              & 0.877           & 0.883       \\
		MSNBC            & 0.910              & 0.910           & 0.910       \\
		Dbpedia          & 0.847              & 0.842           & 0.845       \\
		RSS500           & 0.823              & 0.823           & 0.823       \\
		KORE50           & 0.787              & 0.787           & 0.787       \\
		Micro14        & 0.820              & 0.806           & 0.813       \\
		AQUAINT          & 0.882              & 0.875           & 0.879\\
		\bottomrule
	\end{tabular}
\end{table}

We compare the disambiguation performance of the best setting of Pair-Linking (the one employs the combined NJS\&EES coherence measure) with other state-of-the-art disambiguation systems described as follows:

\begin{itemize}
	\item \textbf{PBoH}~\cite{GaneaGLEH16} is a light-weight system which is based on probabilistic graphical model and \textit{loopy belief propagation} to perform collective disambiguation. The model utilizes Wikipedia statistics about the co-occurrence of words and entities to compute the local matching and pairwise coherence scores.
	
	\item \textbf{DoSeR}~\cite{ZwicklbauerSG16} carefully designs the collective disambiguation algorithm using Personalized PageRank on the mention-candidate graph. The edges are weighted by the cosine similarity between the context and entity embeddings. DoSeR heavily relies on the collective linking algorithm to produce good results.
\end{itemize}

Additionally, we report the results of two simple baselines. One is the prior probability model $P(e|m)$. It simply disambiguates a mention based on the statistics from Wikipedia hyperlinks. The other baseline is the learning to rank Gradient Boosting Tree (GBT)  model which uses only the local confidence score for ranking and selecting candidates. In both baselines, each mention is disambiguated in isolation with other mentions. Therefore, the two can be viewed as local (non-collective) disambiguation models.

Pair-Linking's performance is detailed in Table~\ref{tb:prf1_bestab} and the comparison with other systems are shown in Table~\ref{tb:PL_with_baselines}.
Note that some results of DoSeR and PBoH are slightly different from the ones reported in their original papers~\cite{ZwicklbauerSG16, GaneaGLEH16}.  The reason is that Gerbil (Version 1.2.4) has improved the entity matching and entity validation procedures to adapt to the knowledge base's changes over time.\footnote{http://svn.aksw.org/papers/2016/ISWC\_Gerbil\_Update/public.pdf}

Pair-Linking performs quite well on short text, \ie RSS500, KORE50, Micro14. For the most challenging dataset KORE50, Pair-Linking improves the disambiguation performance by 0.30 F1 compared to the local approach $P(e|m)$ which disambiguates mentions based on the local context. Furthermore, Pair-Linking also outperforms PBoH by 0.14 F1 score on the same dataset. 
Overall, Pair-Linking outperforms the second best disambiguator DoSeR by a large margin (0.045 F1 score).

\subsubsection{Discussion about the NIL mention.}

\begin{table}
	\centering
	\caption{Micro-averaged $F1$ performance of Pair-Linking (with NJS\&EES as coherence measure) with four different percentage of NIL-mention settings. The $F1$ score is calculated on the linkable mentions.}
	\label{tb:robustness_test}
	\begin{tabular}{l|cccc}
		\toprule		
		\textbf{Dataset}                 & \textbf{0\%} & \textbf{20\%} & \textbf{40\%} & \textbf{60\%} \\
		\midrule
		Reuters128&0.859&0.842&0.850&0.848\\
		ACE2004&0.883&0.879&0.900&0.869\\
		MSNBC&0.910&0.890&0.887&0.893\\
		AQUAINT&0.879&0.873&0.875&0.863\\
		
		\bottomrule
	\end{tabular}
\end{table}

In this work, we do not consider the case where a mention refers to a not-in-link (NIL) entity (\ie the entity that does not present in the given knowledge base). One possible solution to detect the NIL mention is to base on the local confidence score. Specifically, a mention is mapped to a NIL entity if the highest local confidence score among its candidates is less than a predefined threshold. However, since the performance of  the threshold-based approach relies on the local confidence modeling which is not the focus of our work, we do not study the NIL detection in this paper. Instead, we will address a more interesting research question: \textit{``How robust is Pair-Linking if NIL mentions are presenting in a document?''}.

Specifically, for each document, we randomly sample few mentions and remove the ground-truth entities  from their candidate sets. We report the disambiguation performance of Pair-Linking with the new setting. Note that in this experiment, we only consider medium-to-long text document which contains sufficient number of mentions and the performance is measured only on the linkable mentions. As reported in Table~\ref{tb:robustness_test}, the presence of NIL mentions does not degrade the performance of Pair-Linking on other linkable mentions, even in the case that $60\%$ of the mentions are NIL.The robust disambiguation performance of Pair-Linking can be explained as follows. Since the local confidence of a NIL-mention and its candidate is usually low, any pair of linking assignment involving the NIL-mention will have low confidence score. As a result, the pair will be selected at the latest in the procedure of Pair-Linking (see Section~\ref{sec:pairlink}). Therefore, the assignment of the NIL-mention is not likely to affect the assignments of other mentions.

\section{Conclusions}
\label{sec:conclusion}

In this work, we study the collective entity disambiguation problem. While conventional approaches assume that all entities mentioned in a document should be densely related, our study reveals the low degree of coherence is not occasional in general text (news, tweet, RSS).

We propose MINTREE, a new tree-based collective linking model that utilizes the weight of minimum spanning tree to measure the coherence in an entity graph. Using the tree-based objective allows us to model the sparse and noisy context effectively. Furthermore, we also show that MINTREE is highly correlated to previously introduced collective linking models, therefore it can be used as a replacement.

Finally, we introduce Pair-Linking, an approximate solution for the MINTREE optimization problem. Despite being simple,  Pair-Linking performs notably fast and achieves comparable accuracy in comparison to other collective linking algorithms. 

\section*{Acknowledgments}
This work was supported by Singapore Ministry of Education Research Fund MOE2014-T2-2-066.

\bibliographystyle{IEEEtran}
\bibliography{arxiv_MINTREE}

\begin{thebibliography}{10}
\providecommand{\url}[1]{#1}
\csname url@samestyle\endcsname
\providecommand{\newblock}{\relax}
\providecommand{\bibinfo}[2]{#2}
\providecommand{\BIBentrySTDinterwordspacing}{\spaceskip=0pt\relax}
\providecommand{\BIBentryALTinterwordstretchfactor}{4}
\providecommand{\BIBentryALTinterwordspacing}{\spaceskip=\fontdimen2\font plus
\BIBentryALTinterwordstretchfactor\fontdimen3\font minus
  \fontdimen4\font\relax}
\providecommand{\BIBforeignlanguage}[2]{{%
\expandafter\ifx\csname l@#1\endcsname\relax
\typeout{** WARNING: IEEEtran.bst: No hyphenation pattern has been}%
\typeout{** loaded for the language `#1'. Using the pattern for}%
\typeout{** the default language instead.}%
\else
\language=\csname l@#1\endcsname
\fi
#2}}
\providecommand{\BIBdecl}{\relax}
\BIBdecl

\bibitem{minh2017neupl}
M.~C. Phan, A.~Sun, Y.~Tay, J.~Han, and C.~Li, ``Neupl: Attention-based
  semantic matching and pair-linking for entity disambiguation,'' in
  \emph{CIKM}, 2017.

\bibitem{PappuBMST17}
A.~Pappu, R.~Blanco, Y.~Mehdad, A.~Stent, and K.~Thadani, ``Lightweight
  multilingual entity extraction and linking,'' in \emph{{WSDM}}, 2017, pp.
  365--374.

\bibitem{GaneaGLEH16}
O.~Ganea, M.~Ganea, A.~Lucchi, C.~Eickhoff, and T.~Hofmann, ``Probabilistic
  bag-of-hyperlinks model for entity linking,'' in \emph{{WWW}}, 2016, pp.
  927--938.

\bibitem{yamada2016joint}
I.~Yamada, H.~Shindo, H.~Takeda, and Y.~Takefuji, ``Joint learning of the
  embedding of words and entities for named entity disambiguation,'' in
  \emph{CoNLL}, 2016.

\bibitem{ShenWH15}
W.~Shen, J.~Wang, and J.~Han, ``Entity linking with a knowledge base: Issues,
  techniques, and solutions,'' \emph{{IEEE} TKDE}, vol.~27, no.~2, pp.
  443--460, 2015.

\bibitem{ShenWLW12}
W.~Shen, J.~Wang, P.~Luo, and M.~Wang, ``{LIEGE:} : link entities in web lists
  with knowledge base,'' in \emph{{SIGKDD}}, 2012, pp. 1424--1432.

\bibitem{GlobersonLCSRP16}
A.~Globerson, N.~Lazic, S.~Chakrabarti, A.~Subramanya, M.~Ringgaard, and
  F.~Pereira, ``Collective entity resolution with multi-focal attention,'' in
  \emph{{ACL}}, 2016.

\bibitem{MurphyWJ99}
K.~P. Murphy, Y.~Weiss, and M.~I. Jordan, ``Loopy belief propagation for
  approximate inference: An empirical study,'' in \emph{{UAI}}, 1999, pp.
  467--475.

\bibitem{RatinovRDA11}
L.~Ratinov, D.~Roth, D.~Downey, and M.~Anderson, ``Local and global algorithms
  for disambiguation to wikipedia,'' in \emph{{ACL}}, 2011, pp. 1375--1384.

\bibitem{FerraginaS10}
P.~Ferragina and U.~Scaiella, ``{TAGME:} on-the-fly annotation of short text
  fragments (by wikipedia entities),'' in \emph{{CIKM}}, 2010, pp. 1625--1628.

\bibitem{austin1991forward}
S.~Austin, R.~Schwartz, and P.~Placeway, ``The forward-backward search
  algorithm,'' in \emph{IEEE ICASSP}, 1991, pp. 697--700.

\bibitem{HoffartYBFPSTTW11}
J.~Hoffart, M.~A. Yosef, I.~Bordino, H.~F{\"{u}}rstenau, M.~Pinkal, M.~Spaniol,
  B.~Taneva, S.~Thater, and G.~Weikum, ``Robust disambiguation of named
  entities in text,'' in \emph{{EMNLP}}, 2011, pp. 782--792.

\bibitem{HanSZ11}
X.~Han, L.~Sun, and J.~Zhao, ``Collective entity linking in web text: a
  graph-based method,'' in \emph{{SIGIR}}, 2011, pp. 765--774.

\bibitem{HacheyRC11}
B.~Hachey, W.~Radford, and J.~R. Curran, ``Graph-based named entity linking
  with wikipedia,'' in \emph{{WISE}}, 2011, pp. 213--226.

\bibitem{GuoB14}
Z.~Guo and D.~Barbosa, ``Robust entity linking via random walks,'' in
  \emph{{CIKM}}, 2014, pp. 499--508.

\bibitem{PiccinnoF14}
F.~Piccinno and P.~Ferragina, ``From tagme to {WAT:} a new entity annotator,''
  in \emph{{ACM} Workshop on Entity Recognition {\&} Disambiguation}, 2014, pp.
  55--62.

\bibitem{AlhelbawyG14}
A.~Alhelbawy and R.~J. Gaizauskas, ``Graph ranking for collective named entity
  disambiguation,'' in \emph{{ACL} Volume 2: Short Papers}, 2014, pp. 75--80.

\bibitem{0001RN14}
A.~Moro, A.~Raganato, and R.~Navigli, ``Entity linking meets word sense
  disambiguation: a unified approach,'' \emph{{TACL}}, vol.~2, pp. 231--244,
  2014.

\bibitem{ZwicklbauerSG16}
S.~Zwicklbauer, C.~Seifert, and M.~Granitzer, ``Robust and collective entity
  disambiguation through semantic embeddings,'' in \emph{{SIGIR}}, 2016, pp.
  425--434.

\bibitem{TangFWZ12}
J.~Tang, A.~C.~M. Fong, B.~Wang, and J.~Zhang, ``A unified probabilistic
  framework for name disambiguation in digital library,'' \emph{TKDE}, vol.~24,
  no.~6, pp. 975--987, 2012.

\bibitem{ZhangH17}
B.~Zhang and M.~A. Hasan, ``Name disambiguation in anonymized graphs using
  network embedding,'' in \emph{CIKM}, 2017, pp. 1239--1248.

\bibitem{CenDSO13}
L.~Cen, E.~C. Dragut, L.~Si, and M.~Ouzzani, ``Author disambiguation by
  hierarchical agglomerative clustering with adaptive stopping criterion,'' in
  \emph{{SIGIR}}, 2013, pp. 741--744.

\bibitem{mikolov2013distributed}
T.~Mikolov, I.~Sutskever, K.~Chen, G.~S. Corrado, and J.~Dean, ``Distributed
  representations of words and phrases and their compositionality,'' in
  \emph{NIPS}, 2013, pp. 3111--3119.

\bibitem{wang2014knowledge}
Z.~Wang, J.~Zhang, J.~Feng, and Z.~Chen, ``Knowledge graph and text jointly
  embedding.'' in \emph{EMNLP}, 2014, pp. 1591--1601.

\bibitem{fang2016entity}
W.~Fang, J.~Zhang, D.~Wang, Z.~Chen, and M.~Li, ``Entity disambiguation by
  knowledge and text jointly embedding,'' in \emph{CoNLL}, 2016.

\bibitem{SunLTYJW15}
Y.~Sun, L.~Lin, D.~Tang, N.~Yang, Z.~Ji, and X.~Wang, ``Modeling mention,
  context and entity with neural networks for entity disambiguation,'' in
  \emph{{IJCAI}}, 2015, pp. 1333--1339.

\bibitem{Francis-LandauD16}
M.~Francis{-}Landau, G.~Durrett, and D.~Klein, ``Capturing semantic similarity
  for entity linking with convolutional neural networks,'' in \emph{{NAACL}
  {HLT}}, 2016, pp. 1256--1261.

\bibitem{MilneW08}
D.~N. Milne and I.~H. Witten, ``Learning to link with wikipedia,'' in
  \emph{{CIKM}}, 2008, pp. 509--518.

\bibitem{GuoCK13}
S.~Guo, M.~Chang, and E.~Kiciman, ``To link or not to link? {A} study on
  end-to-end tweet entity linking,'' in \emph{{HLT-NAACL}}, 2013, pp.
  1020--1030.

\bibitem{kruskal1956shortest}
J.~B. Kruskal, ``On the shortest spanning subtree of a graph and the traveling
  salesman problem,'' \emph{Proceedings of the American Mathematical society},
  vol.~7, no.~1, pp. 48--50, 1956.

\bibitem{prim1957shortest}
R.~C. Prim, ``Shortest connection networks and some generalizations,''
  \emph{Bell Labs Technical Journal}, vol.~36, no.~6, pp. 1389--1401, 1957.

\bibitem{LiSD13}
C.~Li, A.~Sun, and A.~Datta, ``{TSDW:} two-stage word sense disambiguation
  using wikipedia,'' \emph{{JASIST}}, pp. 1203--1223, 2013.

\bibitem{GottipatiJ11}
S.~Gottipati and J.~Jiang, ``Linking entities to a knowledge base with query
  expansion,'' in \emph{{EMNLP}}, 2011, pp. 804--813.

\bibitem{friedman2001greedy}
J.~H. Friedman, ``Greedy function approximation: a gradient boosting machine,''
  \emph{Annals of statistics}, pp. 1189--1232, 2001.

\bibitem{KulkarniSRC09}
S.~Kulkarni, A.~Singh, G.~Ramakrishnan, and S.~Chakrabarti, ``Collective
  annotation of wikipedia entities in web text,'' in \emph{{SIGKDD}}, 2009, pp.
  457--466.

\bibitem{roder2014n3}
M.~R{\"o}der, R.~Usbeck, S.~Hellmann, D.~Gerber, and A.~Both, ``N$^3$-a
  collection of datasets for named entity recognition and disambiguation in the
  nlp interchange format.'' in \emph{LREC}, 2014, pp. 3529--3533.

\bibitem{Cucerzan07}
S.~Cucerzan, ``Large-scale named entity disambiguation based on wikipedia
  data,'' in \emph{EMNLP-CoNLL}, 2007, pp. 708--716.

\bibitem{GerberHBSUN13}
D.~Gerber, S.~Hellmann, L.~B{\"{u}}hmann, T.~Soru, R.~Usbeck, and A.~N. Ngomo,
  ``Real-time {RDF} extraction from unstructured data streams,'' in
  \emph{ISWC}, 2013, pp. 135--150.

\bibitem{HoffartSNTW12}
J.~Hoffart, S.~Seufert, D.~B. Nguyen, M.~Theobald, and G.~Weikum, ``{KORE:}
  keyphrase overlap relatedness for entity disambiguation,'' in \emph{CIKM},
  2012, pp. 545--554.

\bibitem{BasaveRVRSD14}
A.~E.~C. Basave, G.~Rizzo, A.~Varga, M.~Rowe, M.~Stankovic, and A.~Dadzie,
  ``Making sense of microposts ({\#}microposts2014) named entity extraction
  {\&} linking challenge,'' in \emph{WWW}, 2014, pp. 54--60.

\bibitem{UsbeckRNBBBCCCE15}
R.~Usbeck, M.~R{\"{o}}der, A.~N. Ngomo, C.~Baron, A.~Both, M.~Br{\"{u}}mmer,
  D.~Ceccarelli, M.~Cornolti, D.~Cherix, B.~Eickmann, P.~Ferragina, C.~Lemke,
  A.~Moro, R.~Navigli, F.~Piccinno, G.~Rizzo, H.~Sack, R.~Speck, R.~Troncy,
  J.~Waitelonis, and L.~Wesemann, ``{GERBIL:} general entity annotator
  benchmarking framework,'' in \emph{{WWW}}, 2015, pp. 1133--1143.

\end{thebibliography}

\vspace{12mm}
\begin{IEEEbiography}[{\vspace{-2mm}\includegraphics[width=1in,height=1.25in,clip,keepaspectratio]{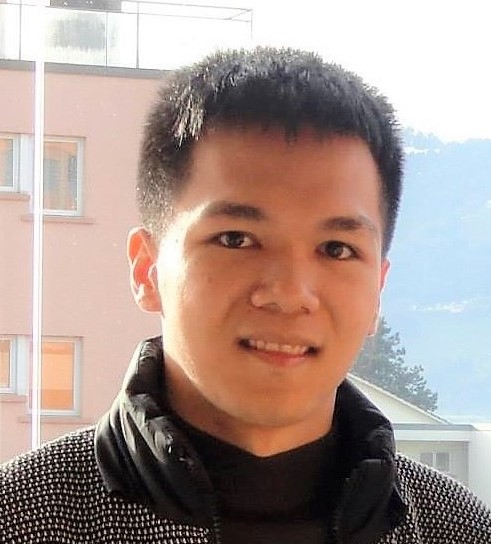}}]{Minh C. Phan}
	is a Ph.D. student at School of Computer Science and Engineering, Nanyang Technological University, under the supervision of Assoc. Prof. Sun Aixin. He obtained the B.E. degree in Computer Science from the same university in 2015. His research interests include information retrieval, text mining, entity resolution and linking.
\end{IEEEbiography}
\vspace{-20mm}
\begin{IEEEbiography}[{\includegraphics[width=1in,height=1.25in,clip,keepaspectratio]{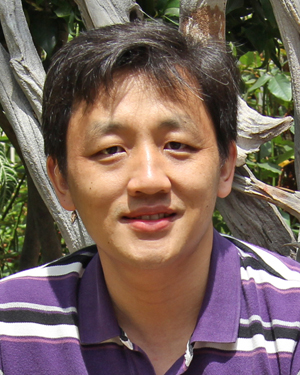}}]{Aixin Sun}
	is an Associate Professor with School of Computer Science and Engineering, Nanyang Technological University, Singapore. He received PhD from the same school in 2004. His research interests include information retrieval, text mining, social computing, and multimedia. His papers appear in major international conferences like SIGIR, KDD, WSDM, ACM Multimedia, and journals including DMKD, TKDE, and JASIST.
\end{IEEEbiography}
\vspace{-20mm}
\begin{IEEEbiography}[{\vspace{-4mm}\includegraphics[width=1in,height=1.25in,clip,keepaspectratio]{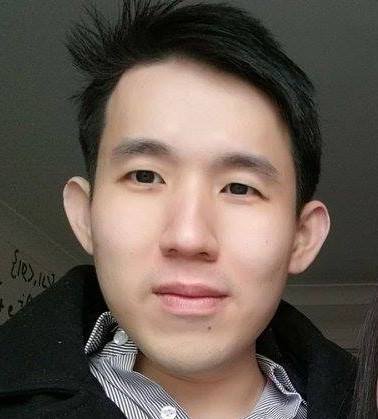}}]{Yi Tay}
	is a Ph.D. student at School of Computer Science and Engineering, Nanyang Technological University, under the supervision of Assoc Prof. Hui Siu Cheung. He receieved the B.E degree in Computer Science from the same university in 2015. His research interests include Deep Learning, NLP, Information Retrieval and Knowledge Graphs.
\end{IEEEbiography}
\vspace{-20mm}
\begin{IEEEbiography}[{\includegraphics[width=1in,height=1.25in,clip,keepaspectratio]{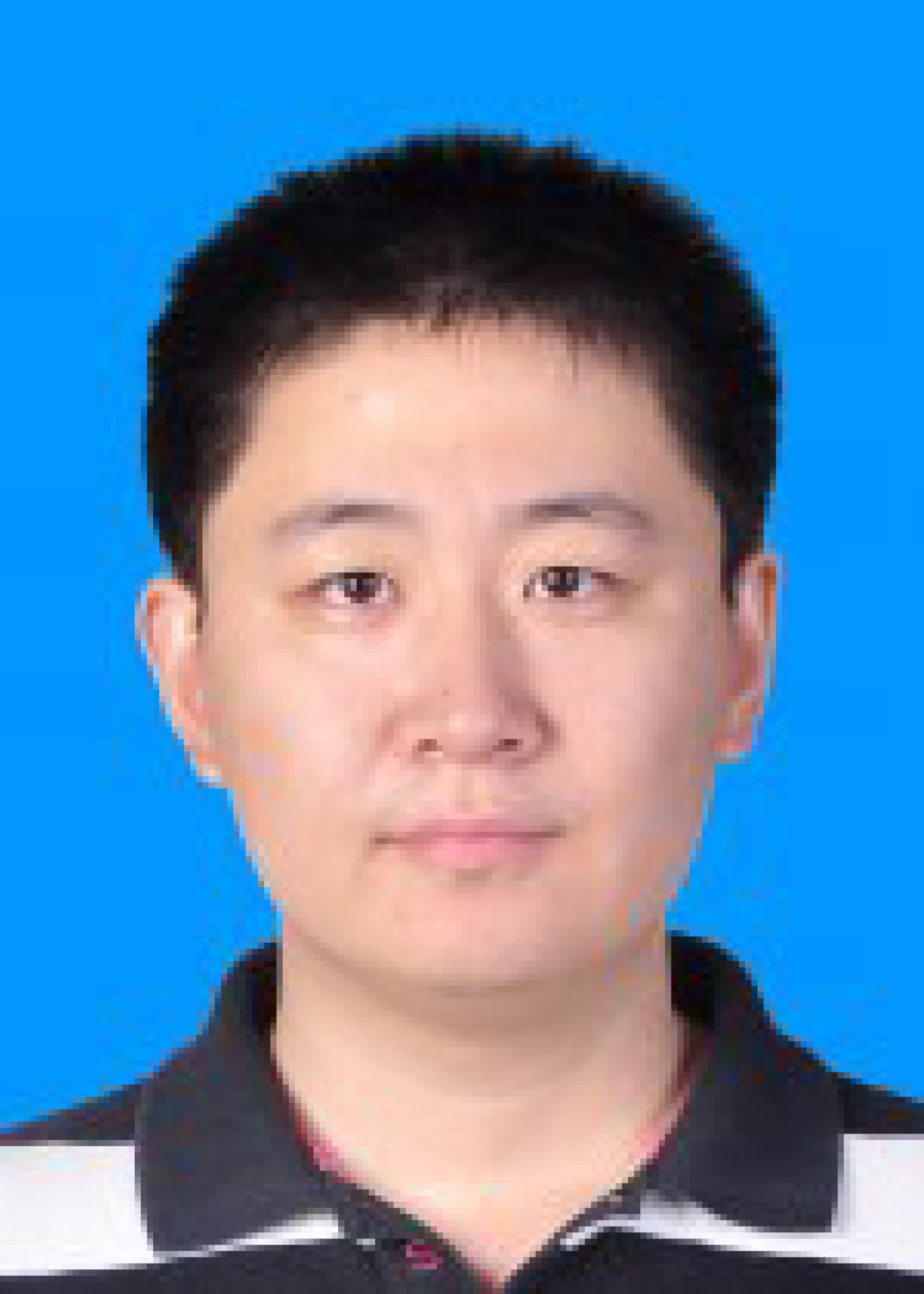}}]{Jialong Han}
	is a postdoctoral research fellow at School of Computer Science and Engineering, Nanyang Technological University. He earned his Ph.D. degree from Renmin University of China in 2015, under the supervision of Prof. Ji-Rong Wen. He obtained his B.E. degree also from Renmin University of China in 2010. His research interests include graph data mining and management, as well as their applications on knowledge graphs.
\end{IEEEbiography}
\vspace{-20mm}
\begin{IEEEbiography}[{\includegraphics[width=1in,height=1.25in,clip,keepaspectratio]{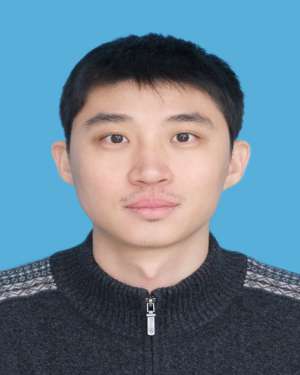}}]{Chenliang Li}
is an Associate Professor at School of Cyber Science and Engineering, Wuhan University, China. He received PhD from Nanyang Technological University, Singapore, in 2013. His research interests include information retrieval, text/web mining, and natural language processing. His papers appear in SIGIR, CIKM, TKDE, TOIS and JASIST.
\end{IEEEbiography}

\end{document}